\documentclass[useAMS,usenatbib]{mn2e}
\usepackage{Times}
\usepackage{psfrag}
\usepackage{pspicture}
\usepackage{graphicx}
\usepackage{amsmath}

\title[LFs from a $\sim$10\,deg$^2$ NB survey]{CF-HiZELS, a $\sim$10\,deg$^2$ emission-line survey with spectroscopic follow-up: H$\alpha$, [O{\sc iii}]+H$\beta$ and [O{\sc ii}] luminosity functions at $\bf z=0.8,1.4$ and $\bf 2.2$}
\author[D. Sobral et al.]{D. Sobral$^{1,2,3}$\thanks{FCT-IF/Veni Fellow. E-mail: sobral@iastro.pt}, J. Matthee$^{3}$, P. N. Best$^{4}$, I. Smail$^{5}$, A. A. Khostovan$^{6}$, B. Milvang-Jensen$^{7}$,\newauthor J. -W. Kim$^{8}$, J. Stott$^{5}$, J. Calhau$^{1,2}$, H. Nayyeri$^{6,9}$, B. Mobasher$^{6}$ \\
$^{1}$ Instituto de Astrof\'{\i}sica e Ci\^{e}ncias do Espa\c{c}o, Universidade de Lisboa, OAL, Tapada da Ajuda, PT1349-018 Lisboa, Portugal \\
$^{2}$ Departamento de F\'{i}sica, Faculdade de Ci\^{e}ncias, Universidade de Lisboa, Edif\'{i}cio C8, Campo Grande, PT1749-016 Lisbon, Portugal \\
$^{3}$ Leiden Observatory, Leiden University, P.O.\ Box 9513, NL-2300 RA Leiden, The Netherlands \\
$^{4}$ SUPA, Institute for Astronomy, Royal Observatory of Edinburgh, Blackford Hill, Edinburgh, EH9 3HJ, UK\\
$^{5}$ Institute of Computational Cosmology, Durham University, South Road, Durham, DH1 3LE, UK\\
$^{6}$ Department of Physics and Astronomy, University of California, 900 University Ave., Riverside, CA 92521, USA \\
$^{7}$ Dark Cosmology Centre, Niels Bohr Institute, University of Copenhagen, Juliane Maries Vej 30, 2100 Copenhagen {\O}, Denmark \\
$^{8}$ Center for the Exploration of the Origin of the Universe, Department of Physics and Astronomy, Seoul National University, Seoul, Korea  \\
$^{9}$ Department of Physics and Astronomy, University of California, Irvine, CA 92697, USA 
}

\begin{document}

\date{Accepted 2015 May 11. Received 2015 May 1; in original form 2015 February 21}

\pagerange{\pageref{firstpage}--\pageref{lastpage}} \pubyear{2015}

\maketitle
\label{firstpage}
\begin{abstract}
We present results from the largest contiguous narrow-band survey in the near-infrared. We have used WIRCam/CFHT and the lowOH2 filter ($1.187\pm0.005$\,$\mu$m) to survey $\approx$10\,deg$^2$ of contiguous extragalactic sky in the SA22 field. A total of $\sim6000$ candidate emission-line galaxies are found. We use deep $ugrizJK$ data to obtain robust photometric redshifts. We combine our data with the High-redshift Emission Line Survey (HiZELS), explore spectroscopic surveys (VVDS, VIPERS) and obtain our own spectroscopic follow-up with KMOS, FMOS and MOSFIRE to derive large samples of high-redshift emission-line selected galaxies: 3471 H$\alpha$ emitters at $z=0.8$, 1343 [O{\sc iii}]+H$\beta$ emitters at $z=1.4$ and 572 [O{\sc ii}] emitters at $z=2.2$. We probe co-moving volumes of $>10^{6}$\,Mpc$^{3}$ and find significant over-densities, including an $8.5$\,$\sigma$ (spectroscopically confirmed) over-density of H$\alpha$ emitters at $z=0.81$. We derive H$\alpha$, [O{\sc iii}]+H$\beta$ and [O{\sc ii}] luminosity functions at $z=0.8,1.4,2.2$, respectively, and present implications for future surveys such as {\it Euclid}. Our uniquely large volumes/areas allow us to sub-divide the samples in thousands of randomised combinations of areas and provide a robust empirical measurement of sample/cosmic variance. We show that surveys for star-forming/emission-line galaxies at a depth similar to ours can only overcome cosmic-variance (errors $<10$\%) if they are based on volumes $>5\times10^{5}$\,Mpc$^{3}$; errors on $L^*$ and $ \phi^*$ due to sample (cosmic) variance on surveys probing $\sim10^4$\,Mpc$^{3}$ and $\sim10^5$\,Mpc$^{3}$ are typically very high: $\sim300$\% and $\sim40-60$\%, respectively.

\end{abstract}


\section{Introduction}

Determining and understanding the star formation history of the Universe is of fundamental importance to improve our understanding of galaxy formation and evolution. Surveys measuring the star formation rate density ($\rho_{\rm SFR}$) as a function of redshift/cosmic time suggest that $\rho_{\rm SFR}$ rises up to $z\sim1-2$ \citep[e.g.][]{Lilly96,Hopkins2006,Karim,Sobral13} and thus reveal that the ``epoch'' of galaxy formation occurs at $z>1$. Most interesting, recent studies are also showing that such behaviour happens for star-forming galaxies at all masses \citep[e.g.][]{Karim,Sobral14}.

There are several star formation indicators that one can use to identify and study star-forming galaxies in cosmological volumes. The most direct tracer of recent star formation is the far ultra-violet (UV) light, coming directly from very massive, short-lived stars. Recombination lines resulting from the strong ionising radiation coming from such stars are also excellent tracers of recent star formation, in particularly H$\alpha$, but forbidden lines such as [O{\sc ii}]$\lambda{3727}$ can also be used. Studies have also used the [O{\sc iii}]$\lambda{5007}$ and other emission lines for this purpose \citep[e.g.][]{Ly2007,Drake}, although with many caveats such as a potential large AGN contamination. Alternative methods for selecting star-forming galaxies are surveying in the far-infrared (FIR) to detect the black-body re-radiation of dust-absorbed UV light from young, massive stars, or surveying in the radio for emission of supernova remnants. There is of course a significant difference between using a star formation indicator to both identify and study star-forming galaxies (which can identify and study star-forming selected galaxies down to some limit), and using such indicators to just measure star formation rates (SFRs) from samples selected by some other means, with more complicated and potentially biased selection functions.

H$\alpha$ stands out as a sensitive star-formation indicator, very well-calibrated and not strongly affected by dust extinction in typical star-forming galaxies, unlike the UV or bluer emission lines \citep[typically A$_{\rm H\alpha}=1$\,mag, e.g.][]{Garn2010a,Sobral12,Ibar13,Stott}. It is also much more sensitive than even the most sensitive FIR or radio surveys, and, with the depths that current instrumentation now allow, H$\alpha$ surveys are able to identify both relatively dust-free and dusty star-forming galaxies, and thus are ideal for an approximately complete selection of star-forming galaxies \citep[see e.g.][]{Oteo15}. When combined with measurements in the FIR, H$\alpha$ becomes an even better SFR selector and indicator \citep[e.g.][]{Garn2010a,Kennicutt09,Ibar13}. Another potential advantage of H$\alpha$ is the possibility to perform surveys with the narrow-band technique, taking advantage of wide-field optical and near-infrared cameras. Narrow-band surveys can probe large areas and, when combined with another narrow-band filter at a close-by wavelength, or a broad-band filter to estimate the continuum, they allow for an effective way to obtain clean, complete samples of emission-line selected galaxies \citep[e.g.][]{Bunker95,Ly2007,G08,Tadaki,Koyama11,CHU11,Lee2012,Sobral12,Sobral13,Koyama13,Stroe14,Stroe14b,An2014}.

The High-redshift(Z) Emission Line Survey \citep[HiZELS;][]{G08,S09a,Sobral12,Sobral13} has exploited various narrow-band filters in the $z$, $J$, $H$ and $K$ bands to undertake deep, wide surveys for line emitters, with a particular strong emphasis on exploring H$\alpha$ emitters across redshift \citep[e.g.][]{Sobral10A,Geach12,Swinbank12,Swinbank12a,Stott,Stott13b,Sobral14}. HiZELS results in a homogeneous selection of H$\alpha$ emitters at $z=0.40$, $z=0.84$, $z=1.47$ and $z=2.23$, over a few $\sim1$\,deg$^2$ areas, detecting around 500-1000 emitters at each redshift down to limiting H$\alpha$ SFRs of $\sim3-10$\,M$_{\odot}$\,yr$^{-1}$ \citep[c.f.][]{Sobral13}. HiZELS allowed for the first fully self-consistent measurement of the evolution of the H$\alpha$ luminosity function from $z=0$ to $z\sim2$, revealing that $\rho_{\rm SFR}$ rises up to $z\sim2$, using one single star formation indicator \citep{Sobral13}. The results also show that the H$\alpha$ star formation history can fully reproduce the evolution of the stellar mass density since $z=2.23$ and that the typical/characteristic SFR of star-forming galaxies (SFR$^*$($z$)) has decreased by a factor $\sim13$ since that time \citep{Sobral14}.

Results from narrow-band and slitless grism surveys are generally in good agreement, but some disagreement by factors of a few have been reported \citep[][]{Ly2007,CHU11,Lee2012,Sobral12,Drake,Colbert,Stroe14}. It is expected that such discrepancies are mostly caused by sample variance (cosmic variance) due to the relatively small areas studied. By probing large enough areas, one can overcome the effect of cosmic variance \citep[e.g.][]{Sobral10A}, and such data can be used to empirically measure the effects of sample variance, instead of having to rely on rather uncertain theoretical/model predictions. Minimising and understanding these discrepancies is not only important for our understanding of galaxy formation and evolution, but also as a forecasting tool for upcoming surveys, such as {\it Euclid} and {\it WFIRST} \citep[e.g.][]{Geach10,Wang13,Colbert}. In particular, determining the bright end of the luminosity function with sufficient accuracy, and minimising cosmic variance, can have a significant impact on estimates of the number of emitters to be recovered with very wide space surveys with e.g. {\it Euclid}.

Narrow-band surveys are also sensitive to many other emission lines besides H$\alpha$ that can be very useful to (potentially) extend measurements to higher redshift, but also to provide alternative/comparative views. These can be used to search for Ly$\alpha$ \citep[e.g.][]{Sobral09b,Matthee14}, but also to trace the evolution of the luminosity function of e.g. oxygen lines, such as [O{\sc ii}] \citep[e.g.][]{Ly2007,Sobral12}. So far studies have mostly just covered relatively small areas \citep[e.g.][]{Ly2007,Drake}, probing down to low luminosities, but at the expense of being strongly affected by cosmic variance, particularly when based on a single field. Significant progress requires probing volumes $\sim10\times$ larger. 
 
In order to address the current shortcomings/limitations, we have undertaken by far the largest narrow-band survey for high redshift emission-line galaxies. Observations were made using the LowOH2 narrow-band filter on CFHT/WIRCam over a $\sim10$\,deg$^2$ area in the SA22 field and we also combine our new data with similar data in the Cosmic Evolution Survey (COSMOS) and the Ultra Deep Survey (UDS) from \cite{Sobral13}. Our study provides a major increase in the sample sizes and statistics, but also probes a new part of the parameter space. The surveyed area corresponds to an area coverage which is $\sim5-10$ times larger than the largest previous emission-line surveys at $z\sim1$ \citep[e.g.][]{S09a,CHU11,Sobral13}. We also undertake a significant spectroscopic follow-up to improve the robustness of our results. Finally, we present the first [O{\sc ii}] luminosity function at $z=2.2$.

The paper is organised in the following way. \S2 outlines the details of the observations, and describes the data reduction, photometric calibration, source extraction and survey limits. \S3 presents the narrow-band selection criteria, the final sample of emitters and the photometric and spectroscopic redshift analysis, including the follow-up observations with MOSFIRE and FMOS. \S3 also presents the selection of different line emitters within the full sample, by using colour-colour selections and photometric redshifts. \S4 presents the methods and corrections applied to the data in order to derive luminosity functions. Results are presented in \S5: the H$\alpha$, H$\beta$+[O{\sc iii}], and [O{\sc ii}] luminosity functions, their evolution to $z\sim2$, implications for {\it Euclid}, and a quantification of sample (cosmic) variance over the full 10\,deg$^2$, which includes the HiZELS data. Finally, \S6 outlines the conclusions. A H$_0=70$\,km\,s$^{-1}$\,Mpc$^{-1}$, $\Omega_M=0.3$ and $\Omega_{\Lambda}=0.7$ cosmology is used. We use a Salpeter IMF and all magnitudes are in the AB system, unless noted otherwise.

%
%
%
\begin{table*}
\begin{center}
\caption{\small{Observation log for the narrow-band observations conducted with the lowOH2 filter on CFHT/WIRCam. A total of 80 pointings, numbered from 0 to 79 (Field ID), were obtained with WIRCam, to cover a total area of approximately 10 deg$^2$. The seeing in all observations was in the range 0.5--0.7$''$ and conditions were photometric.}}
\begin{tabular}{cccccc}
\hline
Field ID & R.A. & Dec.  & Int. time & Dates of observations & 3$\sigma$ limit \\
 & (J2000) & (J2000) & (ks/pixel) &  & (AB, $2''$)  \\ \hline
 0-28 & $22\,18\,00$ to $22\,22\,00$ & $-\,00\,04\,00$ to $+\,00\,06\,24$ &1.0 & 20--30 Sept, 1--18 Oct, 6 Dec 2011, 4 Oct - 3 Nov 2012 & 22.7  \\
 29-53 & $22\,14\,00$ to $22\,18\,00$ & $-\,00\,04\,00$ to $+\,00\,06\,24$ & 1.0 & 4-31 Oct, 1-3 Nov 2012 & 22.7\\ 
 54-79 & $22\,10\,00$ to $22\,14\,00$ & $-\,00\,04\,00$ to $+\,00\,06\,24$ & 1.0 & 20 Sept - 6 Dec 2011, 4 Oct - 3 Nov 2012 & 22.7 \\ \hline
\end{tabular}
\end{center}
\label{tab:observations}
\end{table*}

\section{OBSERVATIONS and DATA REDUCTION}

\subsection{Observations}

\begin{figure}
\centering
\includegraphics[width=8cm]{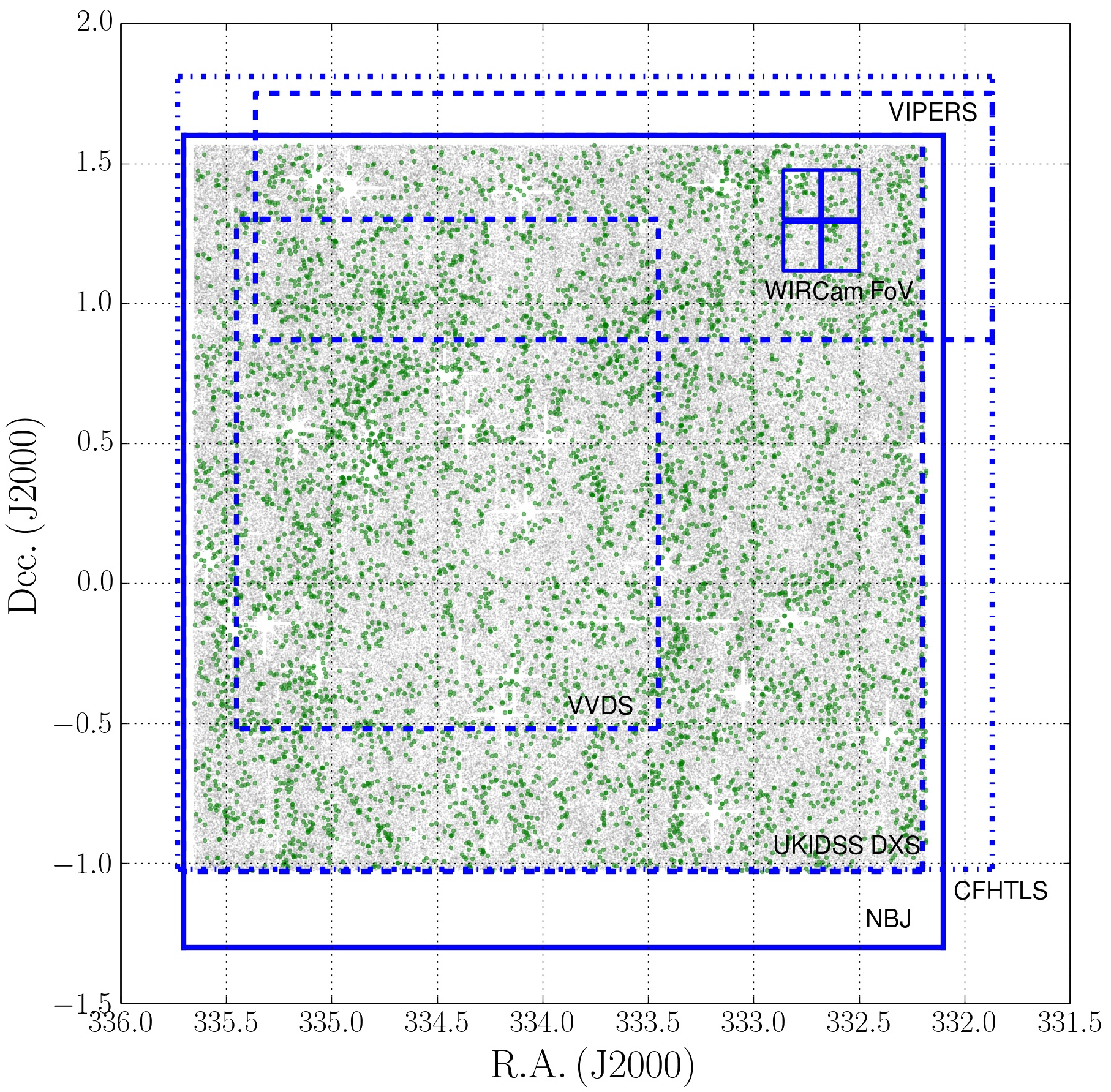}
\caption{\small{Surveyed area in the SA22 with the LowOH2 narrow-band filter (NB$_{\rm J}$) and comparison with other surveys and with the individual WIRCam footprint/field-of-view. For H$\alpha$ emitters at $z = 0.81$, the surveyed area roughly corresponds to a box with $\sim60\times95$ Mpc (physical). The Figure also shows (in grey) all narrow-band detections matched to $J$ and in green the location of all line emitters, irrespectively of redshift (see \S\ref{emitters}). The overlapping regions with CFHTLS W4 ($ugriz$), UKIDSS DXS ($JK$), VVDS and VIPERS ($z_{spect}$) are also shown.}}
\label{fig:surveystrategy}
\end{figure}

Observations were made with the Canada-France-Hawaii Telescope (CFHT) using the Wide-field InfraRed CAMera \citep[WIRCam;][]{Puget2004} during three semesters in queued service mode (Program IDs 2011B/029, 2012A019 and 2012B/016)\footnote{FP7/2007-2013: The research leading to these results has received funding from the European Community's Seventh Framework Programme (FP7/2007-2013) under grant agreement number RG226604 (OPTICON).}. Data were obtained using the LowOH2 narrow-band filter (central wavelength of 1187\,nm\footnote{We fully confirm the central wavelength by exploring all spectroscopic redshifts in our study.} and FWHM, $\Delta \lambda$= 10 nm, also referred in this paper as NB$_J$), and cover a contiguous region of $\sim10$\,deg$^2$ ($\sim3\times3$\,deg$^2$) in the SA22 field (centred at 22:16:00, +00:18:00; see Table \ref{tab:observations}). The SA22 field was chosen for this very wide narrow-band (NB) survey as it presents excellent ancillary data over $\sim9.2$\,deg$^2$ area (see Figure \ref{fig:surveystrategy} for overlap with other surveys).

Observations were conducted from 20 September 2011 to 6 December 2011, and were concluded between 4 October 2012 and 3 November 2012. The seeing over the entire data-set is excellent, in the range $0.5-0.7''$ and observations were done in clear conditions. WIRCam's standard ``paw-print'' configuration of four $2048\times2048$ cryogenically cooled HgCdTe arrays ($0.3045''$/pixel), offset by $\sim1'$ from each other, probe a region of about 0.11\,deg$^2$ (21$'$\,$\times$\,21$'$ field of view at prime focus, see Fig \ref{fig:surveystrategy}) at any given time. In order to cover the full SA22 field, we obtained 80 different pointings. Each pointing was obtained using a dither pattern which resulted in observations being obtained over 10 different positions (individual exposures of 100\,s) per pointing to fill in the detector gaps and minimise the effect of bad pixels and cosmic rays. The final exposure time per pixel is 1.0\,ks. A summary of the observations is given in Table~\ref{tab:observations}.

%
%
%

\subsection{Data reduction}

The narrow-band data were reduced with a dedicated pipeline using {\sc python} and its {\sc pyfits} and {\sc image} modules, based on PfHiZELS \citep[][]{S09a,Sobral13}. Briefly, we start by median combining the dark frames to produce master darks and then use them to dark subtract the individual science frames. We obtain first-pass flat fields by median combining jittered science frames, and use those to flatten the data. We then run {\sc SExtractor} \citep[][]{Bertin1996} on the first-pass flattened frames to produce individual masks. We use those to mask out all individual sources, and, excluding each frame that is being flattened, we produce a final flat field for that frame and flatten the frame. We then use {\sc scamp}\footnote{http://www.astromatic.net/software/scamp} \citep[][]{Bertin2006} to fit a World Coordinate System (WCS) by matching sources detected in individual reduced science images to those in \emph{2MASS} \citep[][]{Skrutskie2006} catalogue. We also use {\sc scamp} to correct for distortions across the field of view by fitting a third order polynomial. Frames are also normalised individually to the same zeropoint by computing the ratio between the expected flux/magnitude from \emph{2MASS} and that found in our data (see more details in Section \ref{extraction}). For both steps we are able to use on average $\sim$ 75 stars per individual frame, resulting in astrometric solutions with typical rms of $\sim0.15''$. Finally, the individual reduced frames from the four detectors are median combined using {\sc swarp}\footnote{http://www.astromatic.net/software/swarp} \citep[][]{Bertin2010}, to obtain stacked reduced data for the entire field. For the broadband $J$ (and $K$) comparison data, we use UKIDSS-DXS-DR10\footnote{http://surveys.roe.ac.uk/wsa/} \citep[][]{Lawrence}.

\subsection{Source Extraction and Survey Limits}
\label{extraction}

We obtain the magnitude zeropoint (ZP) by comparing the magnitudes of the sources in the \emph{2MASS} catalogue to those in our dataset, excluding the faintest ($J>17$, low S/N in \emph{2MASS}) and the brightest ($J<12$, saturated in our data) sources. In order to simplify the analysis, and once accurate ZPs are determined for each stacked image, we scale all images such that ZPs are set to 25, including the broad-band images ($J$, UKIDSS). Sources were extracted using {\sc SExtractor} \citep{Bertin1996} on both narrow-band (NB) and broad-band $J$ (BB) images, using 2$''$ apertures. The 3$\sigma$ AB-magnitude limit for the survey is 22.7 (5$\sigma$: 22.15; c.f. \citealt{Matthee14}), corresponding to an emission line flux limit of $8\times10^{-17}$ erg\,s$^{-1}$\,cm$^{-2}$. This limit is computed by measuring the average background rms of the narrow-band images in 1 million empty 2$''$ diameter apertures and for point-like sources (no aperture correction is applied as it is very small given the excellent seeing, assuming sources are point sources); this is the aperture we use throughout the paper for all measurements and corresponds to about $\sim16$\,kpc physical diameter for the 3 lines investigated here. We note that because we use random aperture measurements, the rms that we measure already accounts for correlations in the noise.

\section{Selection of emitters}\label{emitters}

Down to the 3$\sigma$ limit we detect 346,244 sources in our narrow-band images. Once catalogs with sources in the narrow-band (NB) and in the broad-band (BB) are made, they are matched using a sky algorithm with a maximum separation of 1$''$. Narrow-band sources with no matching broad-band source are likely to be spurious, but they are kept in the catalogue and assigned a $J$ upper limit.

\subsection{Emission line candidates}

In order to robustly select sources that show a real colour excess of the narrow-band over the broad-band, instead of just random scatter or uncertainty in the measurements, two criteria are used. First, the parameter $\Sigma$ \citep[][]{Bunker95} is used to quantify the real excess compared to an excess due to random scatter. This means that the difference between counts in the narrow-band and the broad-band must be higher than the total error times $\Sigma$:
\begin{equation}
c_{\rm NB_J}-c_{\rm BB_J} > \Sigma\delta.
\end{equation}
Here $c_{\rm NB_J}$ and $c_{\rm BB_J}$ are the counts in the narrow- and broad-band respectively, while $\delta$ is the total photometric error, the combination of the errors in both bands:
\begin{equation}
\delta = \sqrt{\pi r_{ap}^2 (\sigma^2_{\rm NB_J}+\sigma^2_{\rm BB_J})},
\end{equation}
where $r_{ap}$ is the radius of the apertures in pixels and $\sigma$ the RMS per pixel in each band. These two equations can be combined into the following equation for $\Sigma$ \citep{Sobral13}:
\begin{equation}
\Sigma=\frac{1-10^{-0.4(\rm BB-NB_J)}}{10^{-0.4(\rm ZP-NB_J)}\sqrt{\pi r^2_{ap} (\sigma^2_{\rm NB_J}+\sigma^2_{\rm BB_J})}}
\end{equation}
ZP is the zeropoint of the NB$_J$ (which is the same as the $BB_J$, because they are both scaled to ZP = 25 in our analysis). We note that the central wavelength of the narrow-band is not perfectly in the centre of the broadband ($J$), but rather at the blue end of the filter, similarly to other narrow-band filters in the $J$ band \citep[see e.g.][]{Sobral13}. Here we correct for this effect using CFHTLS\footnote{Canada-France-Hawaii Telescope Legacy Survey} $z$-band, which is the closest band on the blue side of $J$. Our colour correction (empirical) is given by:
\begin{equation}
BB_J-NB_J = (BB_J-\rm NB_J)_0 + 0.04 (z' -BB_J) - 0.05.
\end{equation}

For sources with no $z$ band ($<3$\,$\sigma$) available (2\%, either because they are too faint in $z$ or because they are masked in CFHTLS), we apply the average correction obtained for all the sources which have reliable $z$ detections ($-0.02$). Thus, we note that this is, on average, a very small correction.

\begin{figure}
\centering
\includegraphics[width=8.5cm]{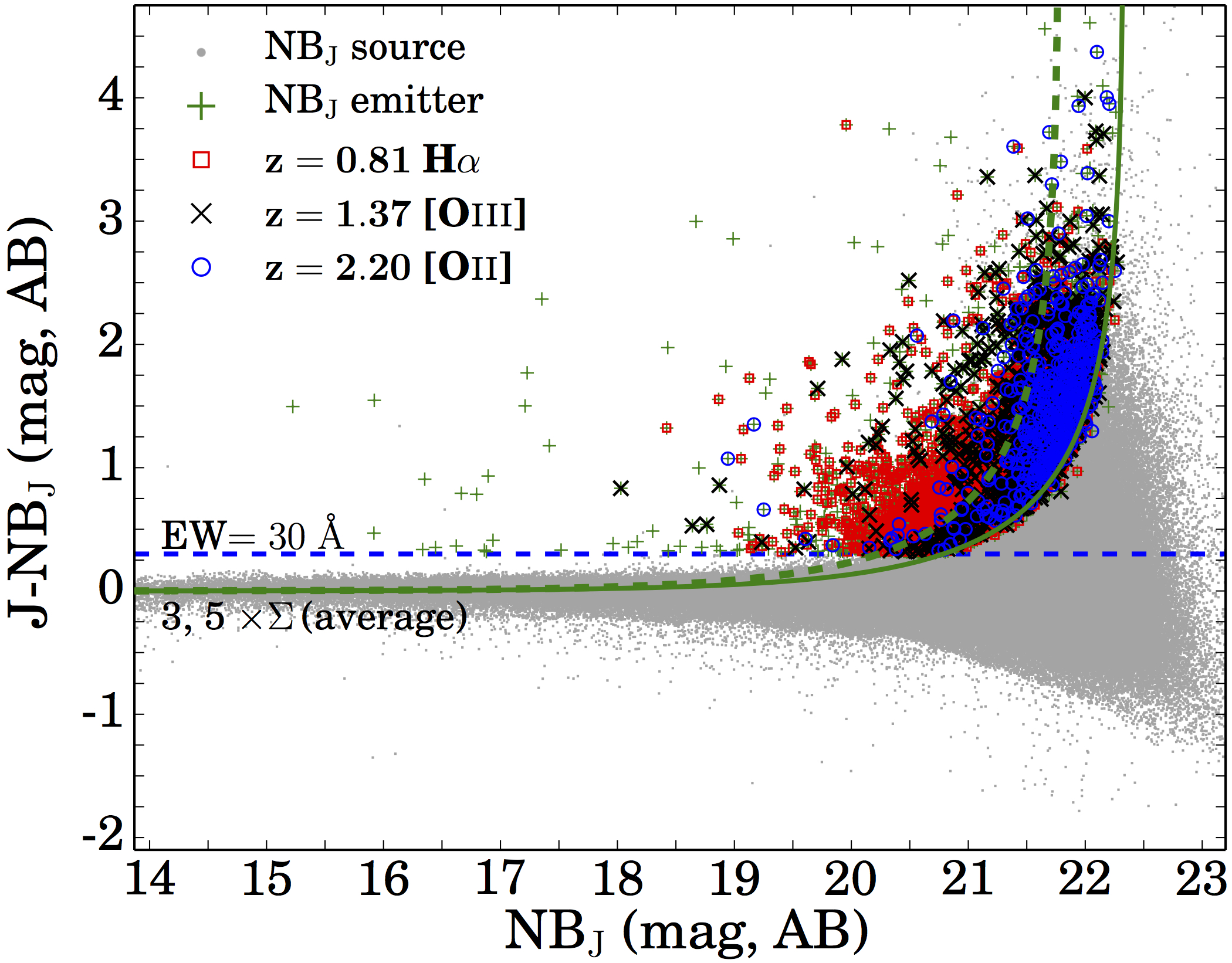}
\caption{\small{Colour-magnitude diagram for our NB$_{\rm J}$ detected sources. The dotted horizontal line is for an observed EW of 30 \AA, which corresponds to J-NB$_J$ $=$ 0.3. The 3 and 5-$\Sigma$ selection curves are shown for the average depth of the survey for reference (but note that some small variations exist across the survey). Candidate line emitters that passed the EW, $\Sigma$ and visual filtering (to exclude spurious sources or artefacts) are shown in colour, while all other non-excess sources are shown in grey. Emitters classed as H$\alpha$ ($z=0.81$), [O{\sc iii}] (+H$\beta$) at $z=1.37$ and [O{\sc ii}] at $z=2.2$, or unclassified, are plotted in different colours as indicated by the key. Note that, as expected, higher redshift emitters are always preferentially found at fainter and fainter magnitudes. Thus, for example, a survey a magnitude shallower would recover little to no [O{\sc ii}] emitters at $z=2.2$.}}
\label{fig:jnbj}
\end{figure}

We classify as potential emitters the sources that have $\Sigma>3$ (see Figure \ref{fig:jnbj}). Table 2 indicates the major emission lines expected. The second criterion for an excess source to be an emitter is that the emission line must have an observed-frame equivalent width (EW, the ratio of the line flux and the continuum flux densities) higher than the scatter at bright magnitudes. This step avoids selecting sources with highly non-uniform continua (with e.g. strong features). We compute EWs by using:
\begin{equation}
EW = \Delta\lambda_{\rm NB_J}\frac{f_{\rm NB_J}-f_{\rm BB_J}}{f_{\rm BB_J}-f_{\rm NB_J}(\Delta\lambda_{\rm NB_J}/\Delta\lambda_{\rm BB_J})}
\end{equation}
$\Delta\lambda_{\rm NB_J}$ and $\Delta\lambda_{\rm BB_J}$ are the widths of the filters and $f_{\rm NBJ}$ and $f_{\rm BB_J}$ are the flux densities for the narrow and broad band respectively. In order to identify a source as a potential line emitter we require it to have EW (observed) higher than 30\,\AA, corresponding to an excess of $\rm J-NB_{\rm J} >0.3$. Note that this will correspond to different rest-frame equivalent widths depending on the line/redshift being looked at. For $z=0.81$, the rest-frame EW limit (H$\alpha$+[N{\sc ii}]) is $\sim17$\,\AA, while for [O{\sc iii}]+H$\beta$, at $z\sim1.4$, and [O{\sc ii}], at $z=2.2$, it is 12.5 and 9 \AA, respectively. These are all relatively low equivalent widths for sources at $z\sim1-2$ \citep[e.g.][]{Fumagalli,Sobral14}, and thus the sample completeness will be high (note that we take this into account when obtaining completeness corrections).

Fluxes of emission lines are calculated as follows:
\begin{equation} 
\label{eq:lineflux}
F_{line} = \Delta\lambda_{NB_J}\frac{f_{NB_J}-f_{BB_J}}{1-(\Delta\lambda_{NB_J}/\Delta\lambda_{BB_J})}.
\end{equation}

Throughout these calculations, the conversion from magnitudes to flux densities ($f_{\nu}$) is:
\begin{equation}
f_{\nu} = 10^{23} \times 10^{-0.4(m_{AB}+48.6)},
\label{eq:abflux}
\end{equation}
where the flux density is given in Jansky and $m_{AB}$ is the magnitude in AB.\\ 

Using our selection criteria, out of the 346,244 NB sources individually detected, 8,599 emitters were selected as potential line emitters. However, many of these are still likely to be artefacts and/or sources in very noisy regions. This is because these numbers do not include any masking for bright stars and their haloes/nor for other artefacts \cite[see e.g.][]{G08, S09a}. We therefore clean our list of potential emitters by visually inspecting all candidates before flagging them as final emitters and produce a final mask. After visually checking all the candidates (obtained before masking), 2,284 are marked as spurious or artefacts \citep[fully consistent with e.g.][]{G08,Sobral13}, with the vast majority being either fake sources/artefacts caused by the many bright ($J<12$) stars in our full coverage, or real sources which have their NB flux boosted (relative to $J$) due to enhanced stripes, again caused by bright stars. We note that the vast majority of these sources could be masked in an automated way, but due to the very wide area of the survey and the different artefacts caused by stars with different bright magnitudes, our method (visually checking every single potential emitter) assured that no region would be overlooked. The remaining sources are excluded because they are detected in low S/N or noisy regions. This leads to a sample of 6,315 potential emitters (see Figure \ref{fig:jnbj}), covering an effective area (after masking) of 7.6\,deg$^2$.


Following \citet{Sobral12}, we then exclude potential stars within the sample by using an optical vs a near-IR colour. In our case, we use $g-z$ versus $J-K$ and identify blue sources in $J-K$ which are very red in $g-z$\footnote{We identify stars as sources with $(g-z)>7\times(J-K)+2$ $\wedge$ $(g-z>1.4)$ $\wedge$ $(J-K)>-0.5$. We check that this procedure eliminates the bulk of spectroscopic stars in both SA22 and COSMOS NB$_J$ samples without excluding real line emitters.}. This results in identifying 339 potential stars, which we remove from the sample, resulting in a final sample of robust line emitters of 5976 line emitters. This density of emitters is in very good agreement with similar surveys of smaller areas \citep[][]{Sobral13}. 

We present the catalogue of emitters containing the final 6315 excess sources, and identify candidate stars with the flag -1 in Appendix \ref{Cats}. The catalogue contains photometry in narrow-band $J$ (from this work, 2$''$) and $J$ (from UKIDSS, obtained by us, also with 2$''$ apertures), line flux, Equivalent Widths and $\Sigma$. Table 3 summarises the catalogue which is available on-line together with the paper.

The data provides the opportunity to identify large statistical samples of emission-line galaxies, mainly H$\alpha$ at $z=0.81$, but also [O{\sc iii}]+H$\beta$ at $z\sim1.3-1.4$, [O{\sc ii}] at $z=2.18$, and Lyman-$\alpha$ candidates at $z=8.8$ \citep[see][]{Matthee14}. Redshifts and probed volumes for the strongest/main emission lines to which our survey is sensitive are found in Table~\ref{redshift}.

%
%
%
%

%
%
%
%
\begin{table}
\centering
\caption{Redshifts at which the LowOH2 filter ($\lambda_c$ = 1187 nm, $\Delta \lambda$ = 10 nm), targets strong emission lines and corresponding probed co-moving volumes per square degree. There are, of course, many other rarer emission lines which the filter can obtain -- see \S3.3. * see \citet{Matthee14} for full details on the search for Ly$\alpha$ at $z=8.8$.}
\begin{tabular}{cccc}
\hline
Emission line & $\lambda_0$ & $z$ & Volume  \\
 & (nm) & & 10$^6$ (Mpc$^3$deg$^{-2}$) \\
\hline
H$\alpha$ & 656.3  & 0.81 $\pm$ 0.01 & 0.11 \\
{[O{\sc iii}]} & 500.7  & 1.37 $\pm$ 0.01 & 0.23 \\
{[O{\sc iii}]} & 495.9  & 1.39 $\pm$ 0.01 & 0.23 \\ 
H$\beta$ &486.1 & 1.44 $\pm$ 0.01 &  0.24 \\
{[O{\sc ii}]} & 372.7  & 2.18 $\pm$ 0.02 & 0.36 \\
Ly$\alpha^*$ & 121.6 & 8.76 $\pm$ 0.04 & 0.52 \\
\hline
\end{tabular}
\label{redshift}
\end{table}

\subsection{Photometry and photometric redshifts}\label{foto}

We combine deep ($\sim23$\,AB) $J$ and $K$ data (UKIDSS DXS DR10) from UKIRT/WFCAM \citep{Lawrence} with CFHT Legacy survey data in $ugriz$ (limit $\sim26$\,AB), to produce a photometric catalogue down to the 5$\sigma$ depth of the $J$ band data ($J$-selected) using 2$''$ apertures. We then match the catalogue with our catalogue of NB$_J$ emitters\footnote{For those without a $J$ band $>5$\,$\sigma$ detection (5\%) we measure their $J$ magnitude centred on the NB$_J$ position and, if undetected above $2\sigma$ (8 sources, 0.13\%), we assign the 2\,$\sigma$ limit.}. By using the $J$-selected catalogue, we measure PSF matched magnitudes in $ugrizK$ (2\,$''$ apertures), and compile a $ugrizJK$ catalogue. We use the photometric catalogue to distinguish between different line emitters by using colour-colour diagnostics, but also by deriving and exploring photometric redshifts.

\begin{figure*}
\centering
\begin{tabular}{cc}
\includegraphics[scale=0.46]{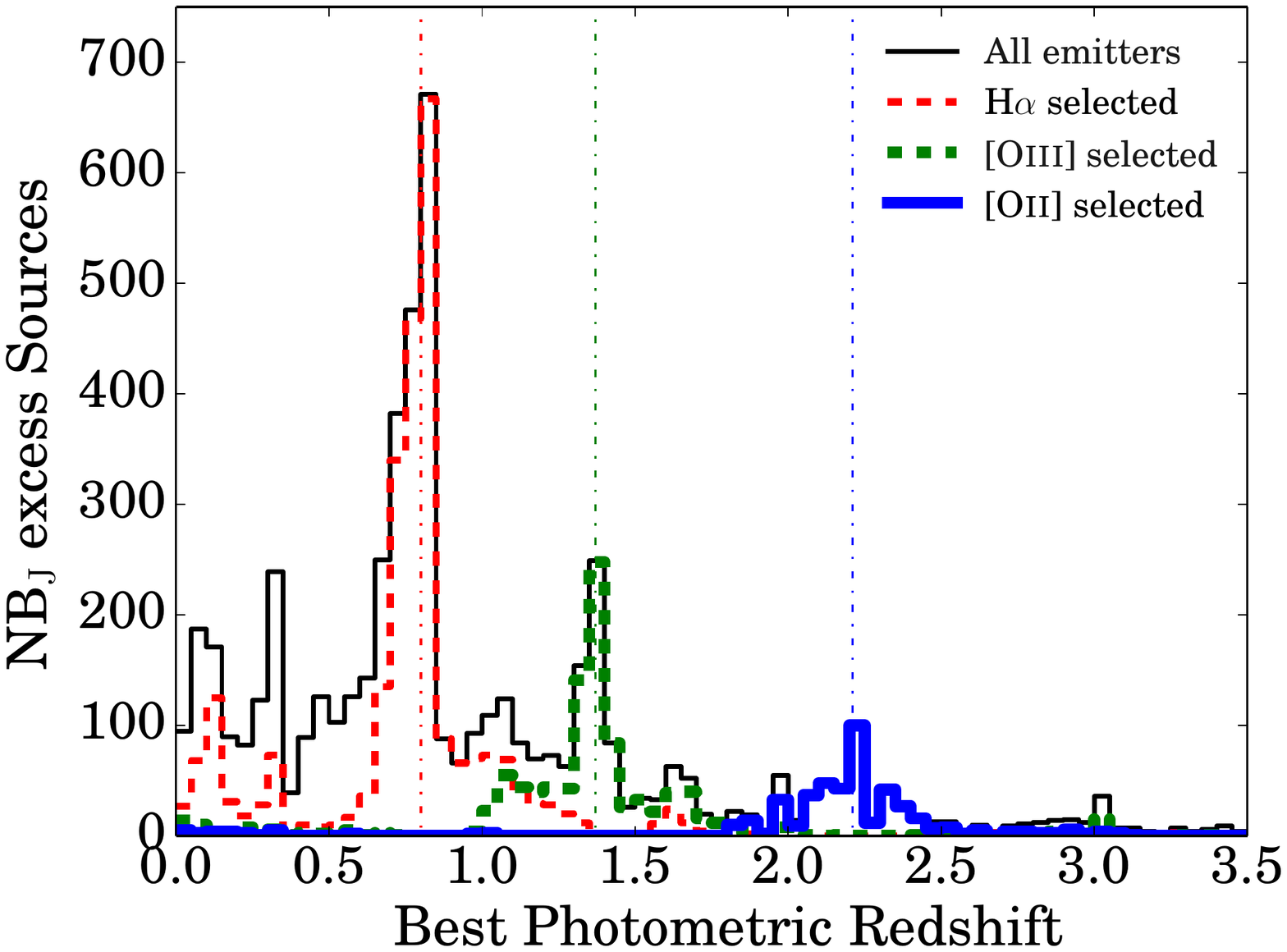}&
\includegraphics[scale=0.46]{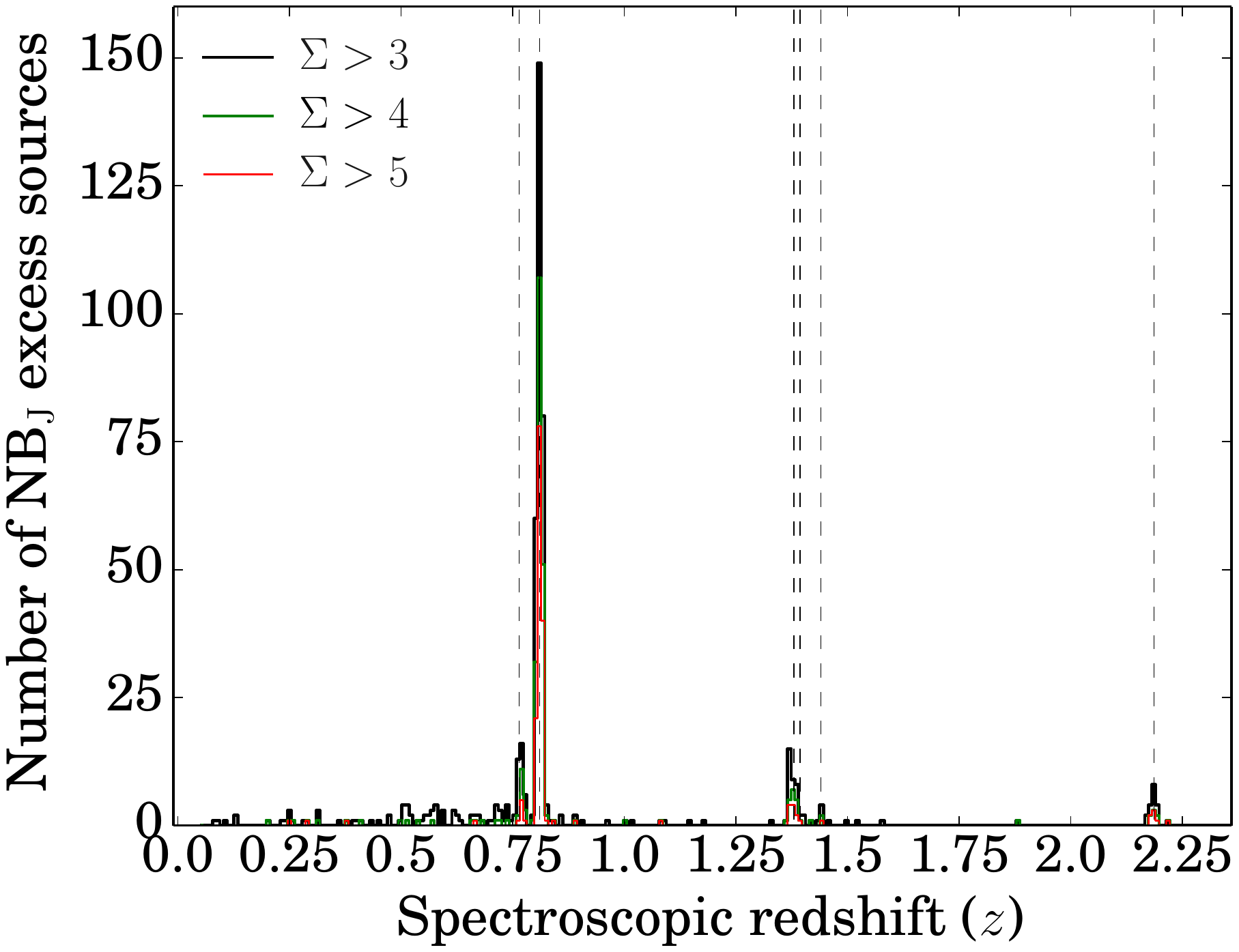}\\
\end{tabular}
\caption{{\it Left:} Photometric redshift distributions obtained with EAZY (including the NB data) for our NB$_J$ selected line emitters in the SA22 field and comparison with the final samples of the different emitters. This shows that the sample of line emitters is dominated by H$\alpha$ emitters at $z=0.8$, followed by H$\beta$+[O{\sc iii}] emitters ($z\sim1.4$), while [O{\sc ii}] emitters at $z\sim2.2$ represent a lower fraction. {\it Right:}The distribution of spectroscopic redshifts of the full sample of line emitters, including our follow-up observations with FMOS on Subaru and MOSFIRE on Keck. This clearly shows that the sample of spectroscopically confirmed line emitters is dominated by H$\alpha$ emitters, followed by [O{\sc iii}], [O{\sc ii}], H$\beta$ and [S{\sc ii}] emitters. Other much rarer lines are found, including Pa$\gamma$10870, N{\sc i}10406,CI9853/9827, [S{\sc iii}]9533, He{\sc ii}8237, O{\sc i}7774 O{\sc ii}7330/7320, [Ar{\sc iii}]7135, [O{\sc i}]6363/6300, He{\sc i}5876, [S{\sc iii}]6311+He{\sc ii}6311. Note that the sample with available spectroscopic redshifts is still dominated by optical follow-up (and observed-optical selection), and thus is highly biased against the higher redshift line emitters. Dashed lines indicate the redshifts of the major emission lines studied in this paper.}
\label{PHOTOZs}
\label{redshift_histo}
\label{PHOTOZs_spec}
\end{figure*}

We compute photometric redshifts by using EAZY \citep{Brammer2008}, which contains a wealth of templates with the main emission lines included. We use $ugrizJK$ photometry, but add our narrow-band photometry as well (we also run EAZY without the NB to control for this addition). In total, we obtain photometric redshifts for 5953 emitters ($>99$\,\%). The photometric redshift distribution of our sample of emitters is shown in Figure \ref{PHOTOZs_spec}. We find that the EAZY photometric redshift distribution has clear peaks at the redshifts of our expected strong emission lines (e.g. H$\alpha$ at $z = 0.8$, H$\beta$/[O{\sc iii}] at $z = 1.4$; see Figure \ref{PHOTOZs_spec}). We also find a tentative peak around $z\sim0.25$, which may be driven by [S{\sc iii}]${9530.6\AA}$ emitters \citep[e.g.][]{Bo2013} -- we confirm some spectroscopically (see Figure \ref{PHOTOZs_spec}). We note that the peaks corresponding to the redshifts of strong emission lines are clearly enhanced by including the narrow-band data. These, probably more accurate, photometric redshifts are used additionally to colour-colour selection and spectroscopic redshifts to distinguish between various emission lines, as outlined in \S3.5. Furthermore, in Figure \ref{fig:redshiftcomparison} we show how our EAZY photometric redshifts (with the inclusion of the NB) compare with all the spectroscopic redshifts available for the sample, revealing good agreement (the spectroscopic redshift distribution of the sample with spectroscopic redshifts is shown in Figure \ref{PHOTOZs_spec}).

%
%
%
%

\subsection{Spectroscopic redshifts: literature compilation}

Spectroscopic redshifts are available from the VIMOS-VLT Deep Survey (VVDS)\footnote{http://cesam.oamp.fr/vvdsproject/index.html; \cite{lefevre13}}, which covered 4\,deg$^2$ in the SA22 field. Additional spectra are available from VIPERS \citep{Garilli13}. In total, 289 of our candidate line emitters have a spectroscopic redshift from either VVDS or VIPERS. However, and particularly for redshifts above 1 (but even for $z\sim0.8$), it becomes significantly more difficult to identify emission lines in the VVDS survey (the survey selection is I$<$22.5 and there are few and weak lines in the optical), as the strongest lines are shifted to the (near)-infrared, or still at bluer wavelengths (like Lyman-$\alpha$). Thus, targeted spectroscopic follow-up (directly in the observed NIR) is ideal to increase the fraction of spectroscopically confirmed line emitters, and allow a more representative evaluation of the range of line emitters within the sample that is not biased towards rest-frame UV bright galaxies.

\subsection{Follow-up observations with MOSFIRE and FMOS}

In order to significantly complement spectroscopic redshifts available from the literature, we follow-up a significant fraction of the sources. We obtained spectra for some of our sources using KMOS \citep[][]{Sharples13}, and these are presented in \cite{Sobral13b} and \cite{Stott13b}. We also followed up some of the bright line emitters with WHT and NTT; these are presented in \cite{Sobral15}.

Furthermore, we have observed large samples of our emitters using FMOS \citep[][]{Kimura10} on Subaru and MOSFIRE \citep[][]{McLean10,McLean12} on Keck. Both instruments are shown to be tremendously efficient for our targets. This is because not only do we know where one of the main emission lines should be found (within a low OH emission window, thus maximising the S/N), but we also know that the vast majority have fluxes high enough to be detected in modest exposure times.

\subsubsection{FMOS observations}

FMOS observations were taken on 15 June 2014, under clear conditions and good seeing (0.7-0.9$''$). We observed sources within the SA22 field in two different configurations, centred on 22 19 57.62 +00 19 35.28 (P1) and 22 19 05.06 +00 52 34.19 (P2). We selected a total of 128 sources from our candidate line emitters in SA22 for P1 and another set of 128 sources from our candidate line emitters in SA22 for P2. The remaining fibres (72) per configuration targeted fillers, calibration stars or were disabled. We used the J-Long ($R=2200$) setting (high resolution mode). Individual exposure times were 900s in each nodding position and we obtained two of each position for a total exposure time of 3.6\,ks per pointing, corresponding to 1.8\,ks of on-target exposure time due to cross-beam switching.

Apart from the raw science frames, we obtained (per configuration) dome-flats and Th-Ar spectral calibration arcs which were used for the reduction. In order to reduce the data, we used the Subaru FMOS reduction pipeline, FMOS Image-Based REduction Package \citep[FIBRE-PAC,][]{Iwamuro2012}. FIBRE-PAC is a combination of IRAF tasks and C programs using the CFITSIO library \citep{Pence1999}. Details are found in \cite{Iwamuro2012}. Briefly, data are flat-fielded and bad pixels removed to begin with. After that, corrections are applied to fix spatial and spectral distortions present and then wavelength calibration is performed. An initial background subtraction is then achieved using the ABAB nodding pattern of the telescope to perform an A-B sky subtraction. Further bad-pixel, detector cross talk, bias difference, distortion, residual background and sky corrections are applied. The 2-d spectra are combined, which in cross beam switching mode means inverting and adding the negative B spectra to the A spectra. The final step is an initial flux calibration, which we cross check with our NB estimated fluxes \citep[see][for more details]{Stott13b}.

%
%
%
\begin{figure*}
\centering
\begin{tabular}{c}
\includegraphics[width=17.cm]{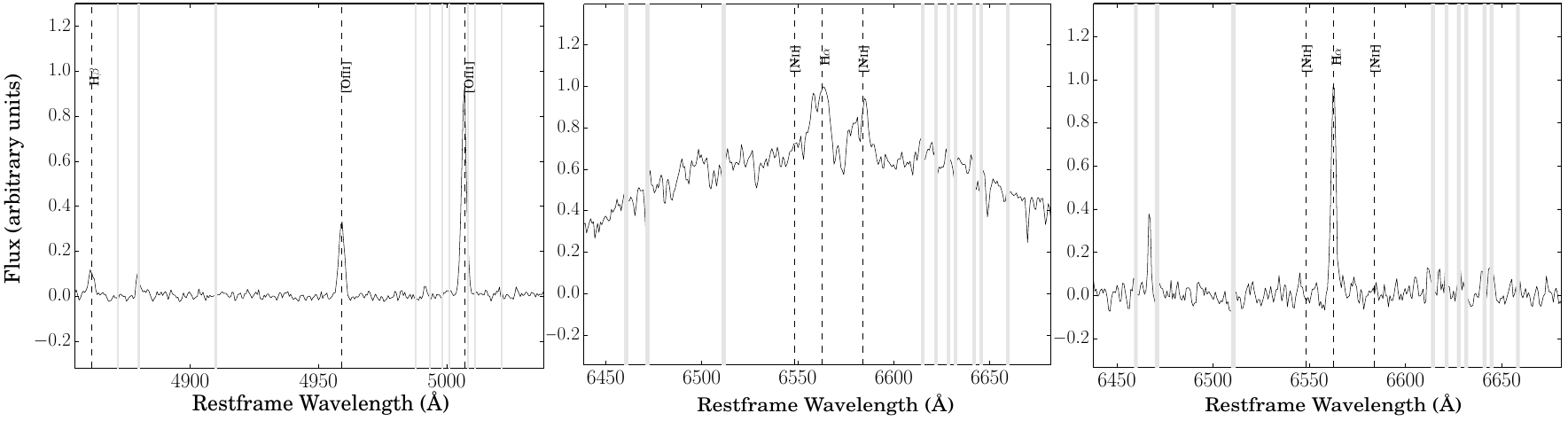}\\
\includegraphics[width=17.cm]{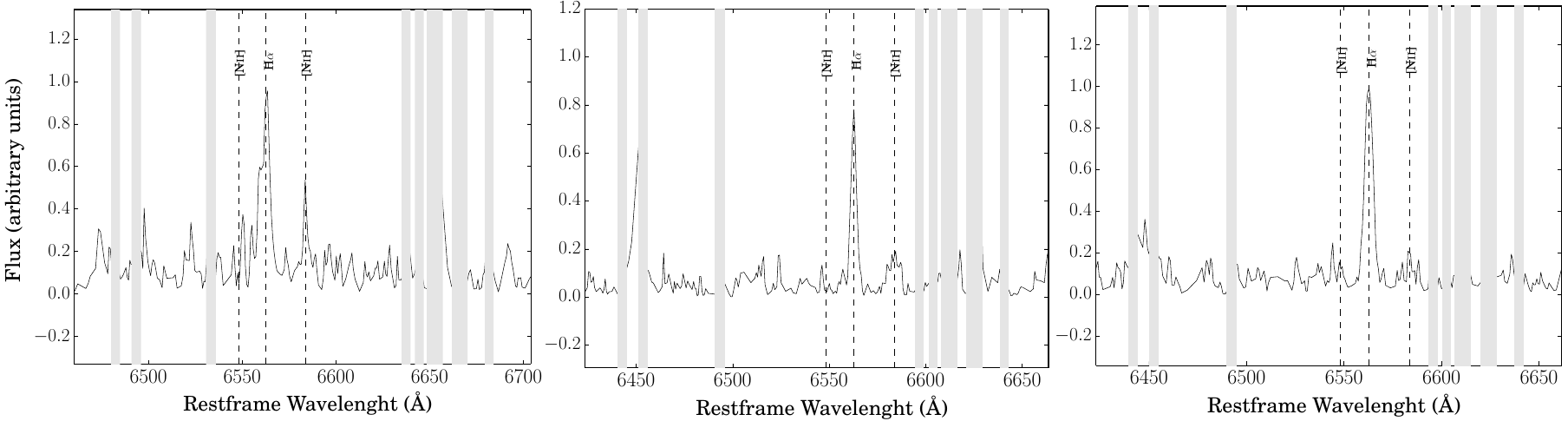}\\
\includegraphics[width=17.cm]{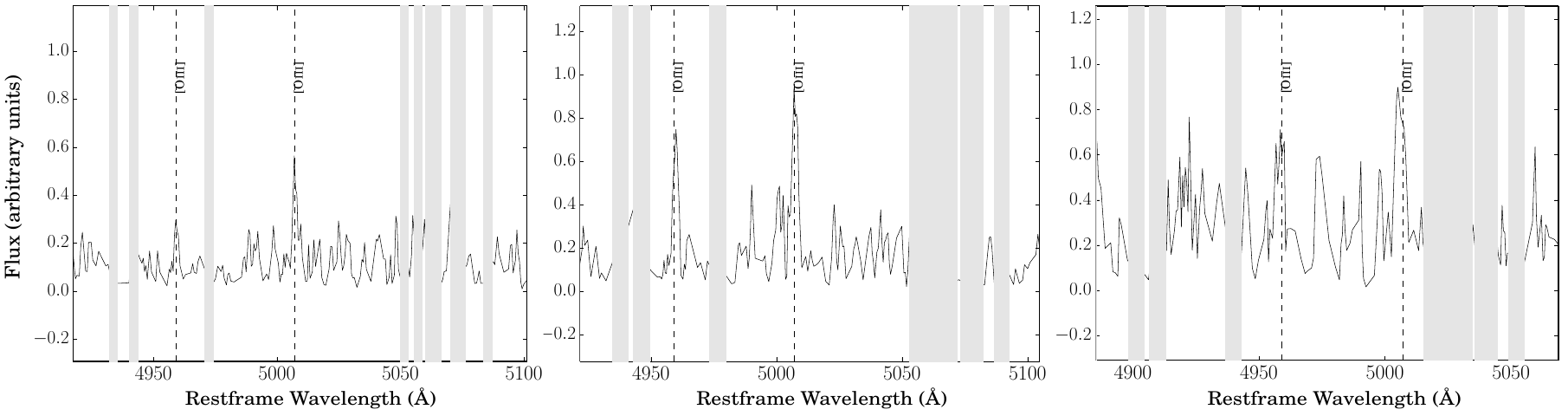}\\
\end{tabular}
\caption{{\it Top:} Some examples of our MOSFIRE spectra. {\it Top left}: one of the confirmed [O{\sc iii}] emitters with a strong detection of H$\beta$, where the emission line being measured and detected by our narrow-band filter is actually [O{\sc iii}]${4959}$. {\it Top middle}: a very luminous broad-line AGN with extremely broad H$\alpha$ line, implying a very massive black hole of $\sim10^9$\,M$_{\odot}$. {\it Top right}: A high EW H$\alpha$ emitter, representative of the less massive population of H$\alpha$ emitters. The locations of the strongest OH lines are indicated as shaded lines, but we note that other weaker OH lines are also present. No smoothing has been applied. {\it Middle:} Representative examples of spectroscopically confirmed H$\alpha$ emitters in our sample within our FMOS data-set. No smoothing has been used. The strongest OH emission lines are represented with shaded regions (but weaker OH lines are not indicated). {\it Bottom:} Representative examples of spectroscopically confirmed [O{\sc iii}] emitters in our sample within our FMOS data-set. No smoothing has been used.}
\label{MOSFIRE_SPECTRA}
\label{EXAMPLES}
\end{figure*}

We obtained redshifts by first identifying the emission line within the NB$_J$ filter profile wavelength range, and then fitting with various combinations. For most spectroscopic redshift determinations there were at least 2 lines, but whenever only one line was present, all solutions were evaluated, and a lower confidence flag was assigned (in almost all cases, such sources were classified as potential H$\alpha$ emitters).

We obtain redshifts for all NB$_J$ emitters with a usable spectrum above a flux of $1.5\times10^{-16}$\,erg\,s$^{-1}$\,cm$^{-2}$, but we also obtain redshifts for more than half of sources with fluxes down to $1.0\times10^{-16}$\,erg\,s$^{-1}$\,cm$^{-2}$. In total, out of the 241 usable spectra, we detect a strong emission line and obtain redshifts for 185 sources. For 135 sources, we are able to obtain a robust spectroscopic redshift, based on at least 2 or more emission lines. Sources without a redshift determination are all lower flux sources, requiring higher exposure times in order to be significantly detected (note that we integrated for only 1.8\.ks per source). We show examples of FMOS spectra in Figure \ref{EXAMPLES}, which shows examples of H$\alpha$ and [O{\sc iii}] emitters.

\subsubsection{MOSFIRE observations}

MOSFIRE observations were obtained on 5 November 2014, under clear conditions. The seeing was 0.8$''$. We observed two masks, both with 0.7$''$ slits. The first one was centred on 22:13:24.89, -00 12 16.31 (p1), while the second one was centred on 22 13 30.27, -00 53 21.24 (p3). Individual exposures were 120\,s in each A and B positions, and we repeated each twice, for a total exposure time of 480\,s per pixel for each of the masks. We reduced the data using the MOSFIRE team data reduction pipeline (DRP\footnote{https://code.google.com/p/mosfire/}). The DRP produces flat-fielded and wavelength calibrated combined 2D spectra for the individual objects. The spectra are wavelength calibrated using the sky lines and then. In practice DRP follows very similar steps to those used to reduce FMOS data. 

For mask p1 we were able to observe 17 of our candidate line emitters, obtaining robust redshifts for all 17 sources. For mask p3 we targeted 8 of our line emitter candidates, and obtained a robust spectroscopic redshift for 7 out of the 8. For the missing source the S/N was too low to detect the emission lines. We note that for most sources the S/N obtained with MOSFIRE are at least comparable, and in most cases much higher than FMOS, despite MOSFIRE data having an exposure time of only $\sim13$ per cent that of FMOS. However, FMOS has a larger multiplexing, allowing to target about 7-10 times more sources (taking into account cross beam switching), and thus it is competitive with MOSFIRE for our targets, i.e., for sources spread over relatively wide areas and that have emission lines which avoid strong OH lines.

In total, we obtained redshifts for 24 sources with MOSFIRE. We show examples of MOSFIRE spectra in Figure \ref{EXAMPLES}. Figure \ref{PHOTOZs_spec} shows the full spectroscopic redshift distribution. 

Together, the FMOS and MOSFIRE data-sets significantly add to the number of spectroscopic redshifts in the sample of line emitters, not only allowing for a spectroscopically confirmed sample of just over 300 H$\alpha$ emitters at $z=0.81$, but also allowing to directly investigate the contamination by the adjacent [N{\sc ii}] line and contributing to a robust correction (see \S\ref{EW_CorrectionNII}). 

\begin{figure}
\centering
\includegraphics[width=8cm]{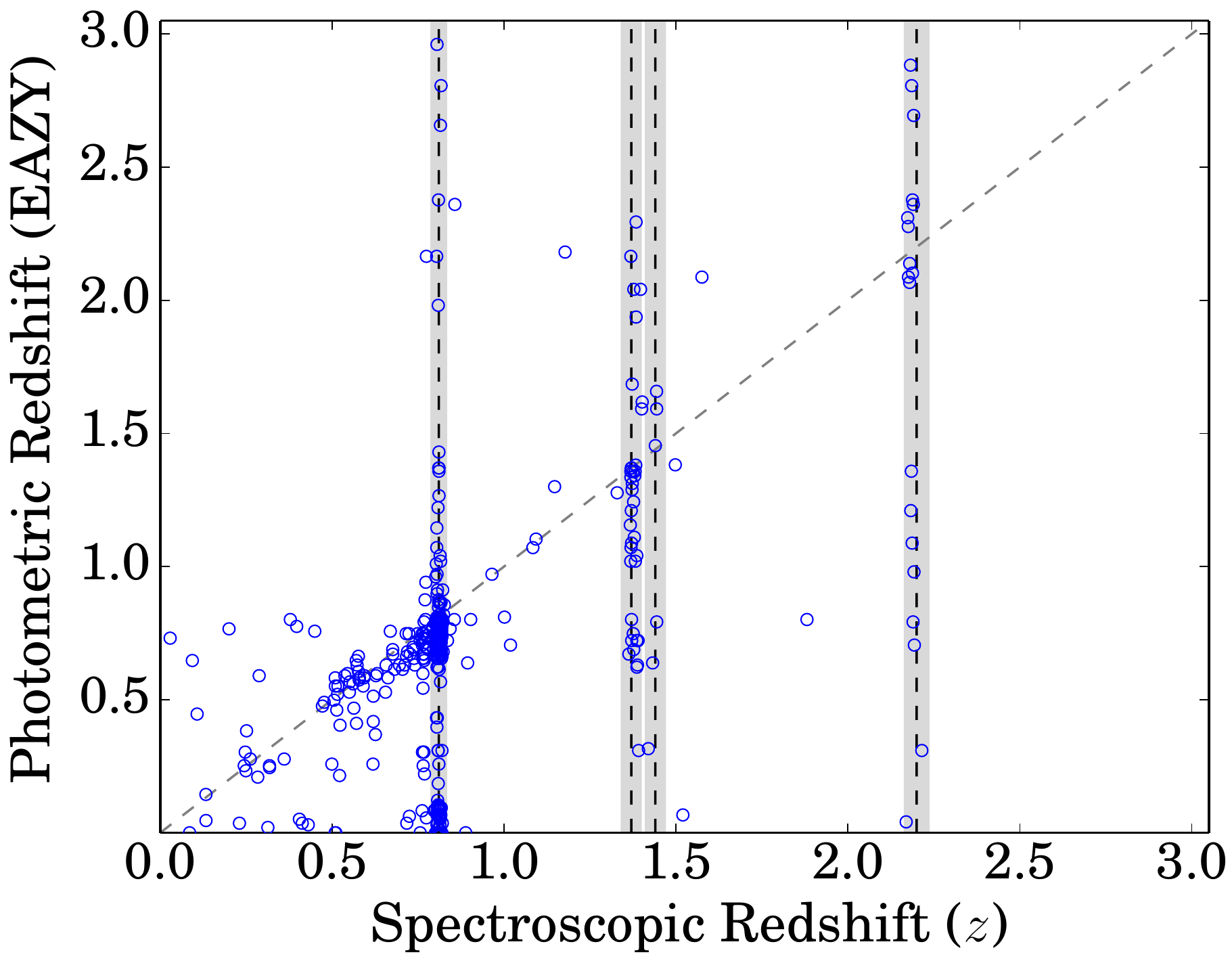}
\caption{\small{Comparison between photometric and spectroscopic redshifts for NB$_J$ selected line emitters for our SA22 NB survey. Redshifts for the emission lines that we select are shown as vertical lines. Many other lines are not labelled, but found, e.g. from low to high redshift: Pa$\gamma$10870, NI10406, CI9853/9827, [SIII]9533, HeII8237, OI7774 OII7330/7320, [ArIII]7135, [SII]6731/6716, H$\alpha$6563, [OI]6363/6300, HeI5876, [SIII]6311+HeII6311, [O{\sc iii}]5007/4959, H$\beta$4861,[O{\sc ii}]3727.}}
\label{fig:redshiftcomparison}
\end{figure}

Finally, by compiling all the 511 spectroscopic redshifts, we test the accuracy of our photometric redshifts for line emitters, which can be seen in Figure $\ref{fig:redshiftcomparison}$. Figure $\ref{fig:redshiftcomparison}$ shows that the redshifts overall agree well, but that there is some scatter, particularly for $z>1$. We use our results, and our full spectroscopic sample, to optimise our selection of different line emitters and to estimate the completeness and contamination of each sample.

%
%
%
%

\subsection{Testing SA22 photometric redshift selection using COSMOS }

In order to further check the validity/accuracy of using our EAZY photometric redshifts ($pz$) based on $ugrizJK$ and our NB for the bulk of our sample, we make use of similar data available in the COSMOS field \citep{Sobral13}. As the COSMOS field \citep[][]{Scoville,Capak,McCracken2012} has been widely studied, there is data available in 30 bands \citep[see e.g.][]{Ilbert09,Ilbert10,Ilbert}, making photometric redshifts much more reliable, while there are also relatively more spectroscopic redshifts available in the literature. Thus, in order to have a fully comparable sample, we compute photometric redshifts in COSMOS by following the same method as for our SA22 sample, i.e., by using $ugrizJK$ and NB$_J$. This allows us to directly compare photometric redshifts for a sample of similar line emitters and investigate any biases/incompleteness due to the limited availability of photometric bands.

\begin{figure}
\centering
\includegraphics[width=8.5cm]{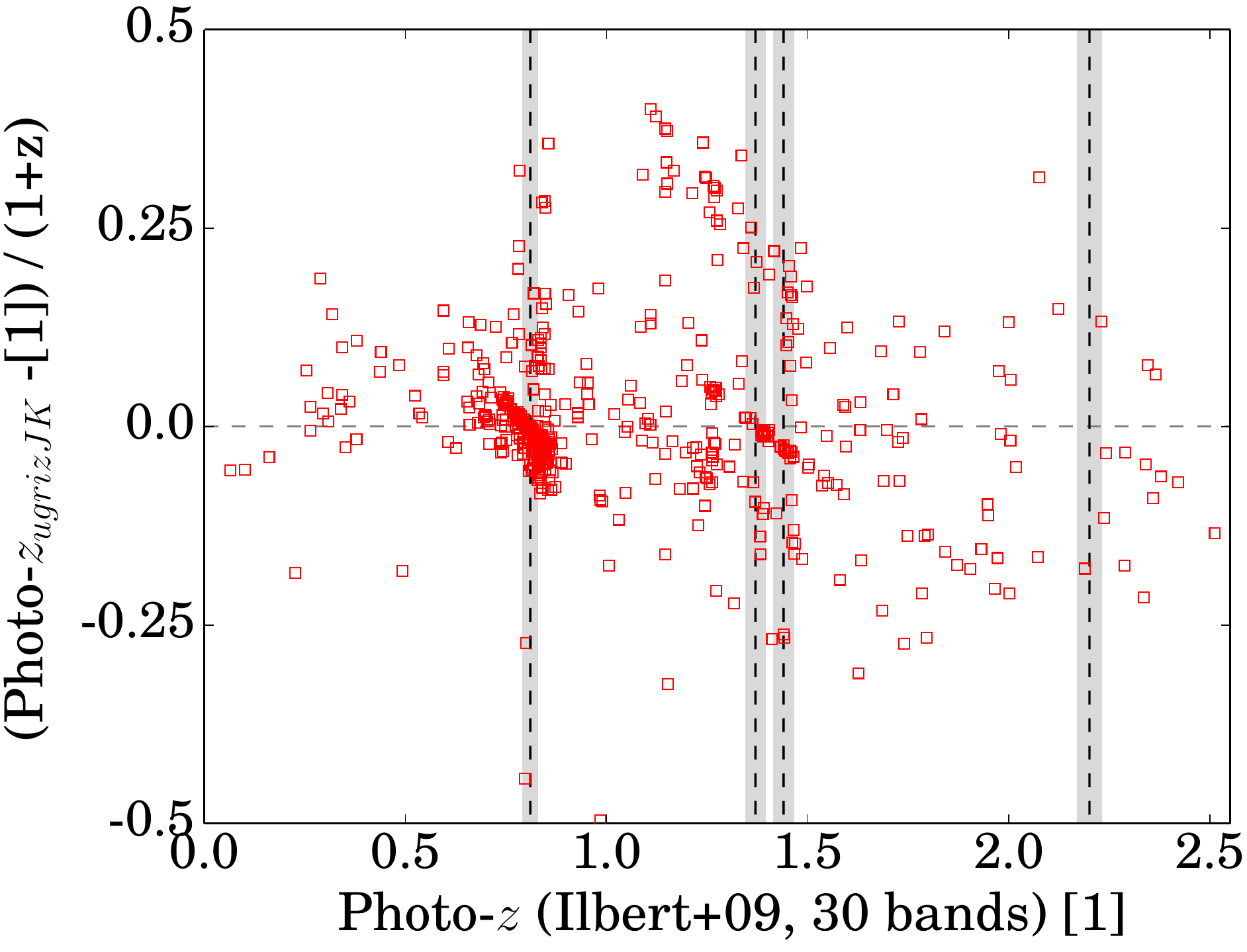}
\caption{A comparison between photometric redshifts for the COSMOS sample of NB$_J$ line emitters (from \citet{Sobral13}), derived by us with EAZY using $ugrizJK$ and NB magnitudes and those derived by \citet{Ilbert09} with 30 bands from FUV to IRAC 8\,$\mu$m. We find a good correlation between the two estimates, with a relatively small scatter below $z<1$. The scatter increases at higher redshift, particularly due to the lack of IRAC bands, but our photometric redshift ranges used for selection, together with our colour-colour selections are able to tackle that.}
\label{fig:photozs}
\end{figure}

For the 700 sources selected as line-emitters in COSMOS by \citet{Sobral13} in their NB$_J$ band, 76 had spectroscopic redshifts \citep{Lilly2009}. As a first step, we derive photometric redshifts using $ugrizJK$ in COSMOS without the inclusion of the NB$_{\rm J}$ filter and compare them with those derived with the inclusion of the NB. We find that both photometric redshift sets are in good agreement for most redshifts, except for the redshifts of strong emission lines such as H$\alpha$ at $z = 0.8$; for these photometric redshifts clearly benefit from the addition of the NB and are always more accurate at recovering spectroscopic redshifts, resulting in typical photometric redshift uncertainties of $\approx0.05$. We compare the photometric redshifts obtained in COSMOS, with all bands, from \cite{Ilbert09}, with those we derive with EAZY using the restricted set of bands including NB. Our results are shown in Figure \ref{fig:photozs}. We find good agreement between both. We find that for the bright ($NB_J < 21.2$, AB) subset of the sample, the EAZY photometric redshifts are consistent to $\approx0.02$. For fainter sources, the EAZY photometric redshifts can sometimes differ significantly from those using all bands, due to larger error-bars for the faint sources, but also because the lack of bands redder than $K$ and the lack of $H$ and UV bands.

%
%
%

Finally, using the data and computed photometric redshifts in COSMOS we check our selection of emitters that we will apply for SA22. We apply the same selection criteria (see \S\ref{selection}) in COSMOS, using both the \cite{Ilbert09} photo-$z$s based on 30-bands and our photometric redshifts with $ugrizJK$+NB. We find that our SA22-like selection criteria (photo-$z$ with just $ugrizJK$+NB and colour-colour selections; see Section \ref{selection}) applied to COSMOS recovers 90\% of the H$\alpha$ emitters (97 \% of the spectroscopically confirmed H$\alpha$ emitters) that are found with much more accurate photometric redshifts. Our photometric redshifts tend to sometimes favour higher redshift solutions compared to \cite{Ilbert09}, which leads to 1 spectroscopic H$\alpha$ emitter being selected as [O{\sc ii}] emitter (but this is a small effect). However, apart from this, the selection is very clean and leads to the conclusion that for our emitters, $ugrizJK$+NB photometric redshifts + colour-colour selections are sufficient and comparable to the best photometric redshifts in COSMOS. We check that similar conclusions are reached for H$\beta$+[O{\sc iii}] and [O{\sc ii}] emitters. Therefore, our samples selected from SA22 and COSMOS (all selected with the same selection criteria) are unlikely to have strong differences and should have very comparable completeness ($\sim90$\%) and contamination ($\sim10$\,\%) fractions.

\subsection{Selection of H$\alpha$, [O{\sc iii}]/H$\beta$ \& [O{\sc ii}] emitters}\label{selection}

The selection of H$\alpha$, [O{\sc iii}]/H$\beta$ \& [O{\sc ii}] emitters at $z=0.81,1.4,2.2$, respectively, is done following \cite{Sobral13}, by using a combination of photometric redshifts (and spectroscopic redshifts, when available) and colour-colour selections optimised for star-forming galaxies at the redshifts of interest ($z\sim0.8,1.4,2.2$). We apply these selections not only to the sample of emitters in SA22 presented here, but also to COSMOS and UDS \citep[from][]{Sobral13}, in order to obtain larger samples over a larger number of independent volumes. Spectroscopic redshifts are also used to evaluate the completeness and contamination of the sample, although due to the selection function of most of the literature spectroscopic redshifts they are only really useful for sources up to $z\sim0.8$. The selection criteria applied here are the same as \cite{Sobral13} for H$\alpha$ at $z\sim0.8$, while for [O{\sc iii}]+H$\beta$ at $z\sim1.4$ we use their criteria to select H$\alpha$ sources at $z\sim1.47$, and for [O{\sc ii}] emitters at $z\sim2.2$ we use their criteria for $z\sim2.2$ H$\alpha$ emitters. We check with COSMOS and UDS that these criteria work well \citep[see][for more details]{Khostovan15}.

\begin{figure}
\centering
\includegraphics[width=8cm]{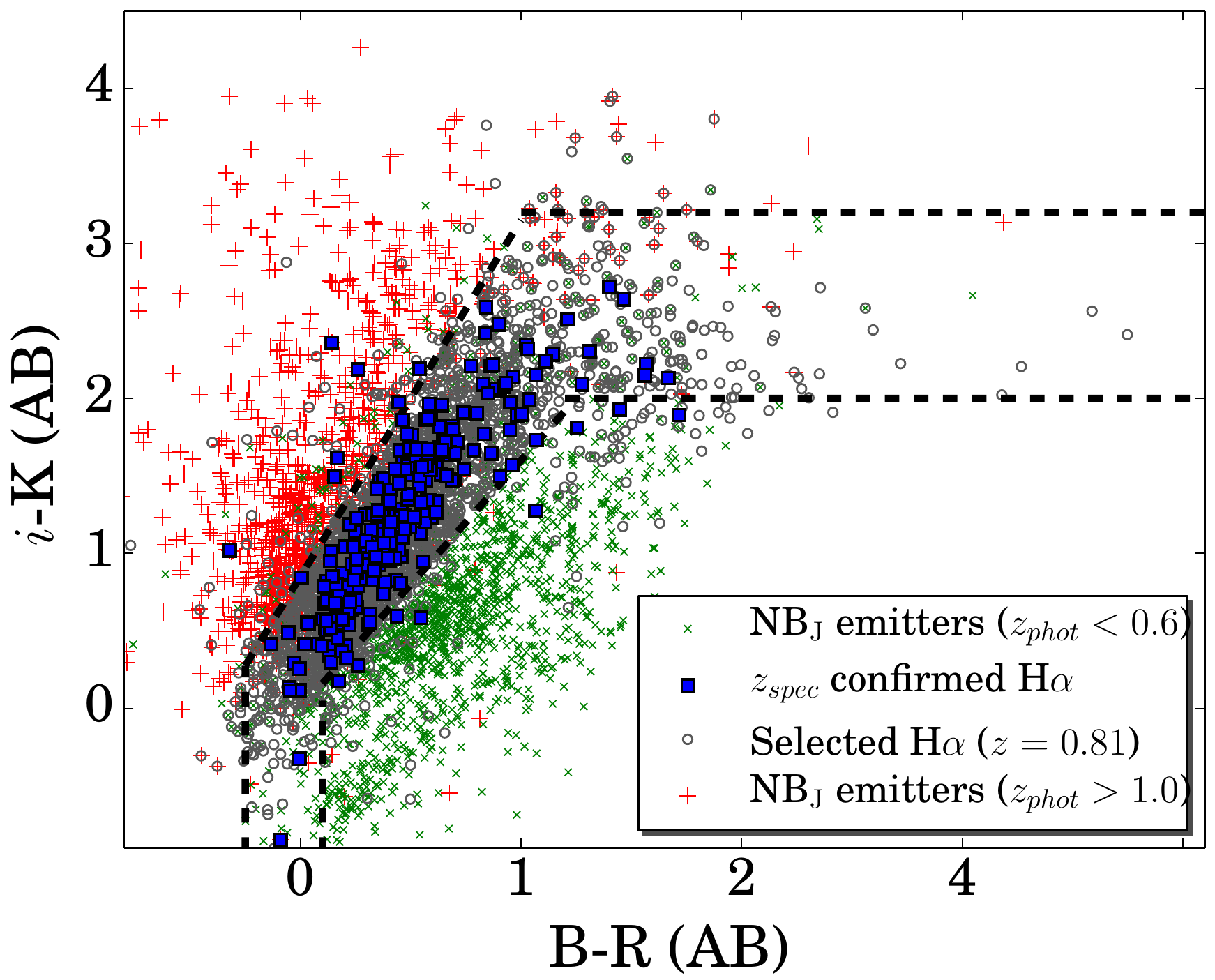}
\caption{\small{$B-R$ vs $i-K$ colour-colour separation \citep{S09a} that we use (together with photometric redshifts) to identify H$\alpha$ emitters within the full sample emitters. Here, only narrow-band line emitters are plotted. Note how the selection is able to distinguish between $z\sim0.8$ sources, those at $z>1$ and those at lower redshifts ($z<0.6$), but also that only $\sim5$\,\% of all spectroscopically confirmed H$\alpha$ emitters do not fall on our BRiK colour-colour selection, thus implying high completeness.}}
\label{fig:colourselection}
\end{figure}

The following Sections describe the specific selection criteria used to identify the different line emitters, but briefly: i) we apply a photometric redshift range, centred on the expected redshift of the line, and that takes into account typical errors (motivated by the estimated uncertainties in the photometric redshifts and on e.g. Figures \ref {fig:redshiftcomparison} and \ref{fig:photozs}) and the proximity of any other strong emission line, ii) we use colour-colour selections (see e.g. Figure \ref{fig:colourselection}), to increase completeness and iii) we explicitly remove any spectroscopically confirmed contaminant and include any spectroscopic confirmed source in the sample. Finally, if a source is classified as e.g. H$\alpha$ it can no longer be selected as a higher redshift emitter and, if classified as [O{\sc iii}]+H$\beta$ it will not be able to be selected as an [O{\sc ii}] emitter. We use our spectroscopic redshifts in order to test different photometric redshift cuts and study the completeness and contamination of our samples. We find that we need to use a wider cut in photometric redshifts than \cite{Sobral13} in order to maximise the completeness (95\% for H$\alpha$ emitters), while maintaining the contamination at a low level of $\sim10-15$\%. This a simple consequence of our photometric redshifts in SA22 having a larger uncertainty.

%
%
%
%

\subsubsection{H$\alpha$ emitters at $z=0.81$}

H$\alpha$ emitters at $z\sim0.8$ can be distinguished from lower and higher redshift emitters very efficiently by using i) our photometric redshifts and ii) the $BRiK$ \citep{S09a} colour-colour diagram (see Figure \ref{fig:colourselection})\footnote{For SA22 we estimate B band photometry by using g and mimicking the necessary empirical correction extracted from the COSMOS and UDS \citep{Sobral13}, for which $u$,$g$ and B data is available. The correction is: $(B-r)_{cor} = (g-r)\times1.1657-0.223$}, in combination with photometric redshifts and spectroscopic redshifts \citep{S09a}. We classify a source as H$\alpha$ if i) its photometric redshift is within $z_{phot}=0.70-0.95$ (1523 sources) or ii) if it satisfies the $BRiK$ colour-colour criteria (Figure \ref{fig:colourselection}) and does not have $1.3<z_{phot}<1.5$ (likely [O{\sc iii}]) or $2.0<z_{phot}<2.4$ (likely [O{\sc ii}]): this step adds 1330 extra sources (a further 200 sources that satisfy the $BRiK$ selection are not included due to their photo-zs clearly pointing towards higher redshift line emitters). We use the $BRiK$ colour-colour selection to guarantee high completeness even for the faintest sources, where photometric redshifts become unreliable and/or unavailable. We recover 286 spectroscopically confirmed H$\alpha$ emitters (see Figure \ref{EXAMPLES} for a few typical examples), remove 34 sources spectroscopically confirmed to be other emission lines either than H$\alpha$ ([S{\sc ii}] emitters are clearly the largest contaminant, but there are also a few low redshift emission lines for which VVDS and VIPERS is extremely complete), and introduce 15 spectroscopically confirmed H$\alpha$ emitters which did not satisfy the photometric selection criteria. Comparing this number with the 286 emitters spectroscopically confirmed to be H$\alpha$ which our selection successfully recovers indicates $\sim$95\% completeness. With the current spectroscopic follow-up, we estimate contamination on the level of $\sim10$\%, being dominated by [S{\sc ii}] emitters at $z=0.76$. We note that this is a consequence of photometric redshifts and colour-colour selections not being able to completely distinguish between H$\alpha$ and [S{\sc ii}] emitters as they are very close in redshift.

In total, we select 2834 H$\alpha$ emitters in SA22 (photometric redshift distribution of the final sample is shown in Figure \ref{PHOTOZs_spec}). Out of all H$\alpha$ emitters in SA22, 295 are spectroscopically confirmed (see Figure \ref{EXAMPLES} for examples from FMOS, and Figure \ref{Stack_spectra} for the composite spectra), from VVDS (75), VIPERS (74), KMOS \citep[20;][]{Sobral13b,Stott14}, 10 sources followed-up with NTT and WHT, 16 with MOSFIRE and 100 with FMOS.

\subsubsection{[OIII] and H$\beta$ emitters at $z\sim1.4$} \label{OIII_Hb}

In order to select [O{\sc iii}]+H$\beta$ emitters\footnote{It is not possible to completely distinguish, with photometric redshifts and/or colour-colour selections between [O{\sc iii}] and H$\beta$ emitters due to a very similar redshift, although we expect [O{\sc iii}] emitters to dominate the sample -- see \S \ref{OIII_Hbeta_fluxes}}, we use the same selection as for H$\alpha$ emitters at $z=1.47$ in \citet{Sobral13}. Our selection criteria are the following: i) we disregarding all candidate H$\alpha$ emitters obtained in the previous subsection (rejection of 2834 sources), ii) we use our photometric redshifts and apply the selection $1.0<z_{phot}<1.8$ (911 sources) and iii) we use our $BzK$ selection to identify $z\sim1.4$ sources from which we remove sources likely to lie at $z>2$ on the basis of their $izK$ colours \citep[see][]{Sobral13} and a further 41 sources with $2.0<z_{phot}<2.4$ which are likely to be [O{\sc ii}] emitters; this adds 141 sources to the sample. Just like for H$\alpha$, we also use spectroscopic redshift information to remove eight non-[O{\sc iii}]+H$\beta$ emitters in the sample and add twelve spectroscopically confirmed [O{\sc iii}]+H$\beta$ emitters. Based on the limited spectroscopic information, we estimate $\sim80\%$ completeness and $\sim15$\% contamination. We have a total of 46 spectroscopically confirmed sources; in Figure \ref{EXAMPLES} we show some examples. By following these selection criteria, we select 1056 [O{\sc iii}]+H$\beta$ emitters in SA22 (photometric redshift distribution of the final sample is shown in Figure \ref{PHOTOZs_spec}). We note that our limited spectroscopic follow-up already allows us to constrain the fraction of H$\beta$ and [O{\sc iii}] emitters within this sample. Within the full sample of spectroscopically confirmed [O{\sc iii}]+H$\beta$ emitters, we find that $\sim16$\, per cent are H$\beta$ emitters (H$\beta$ line detected by the NB filter), and that H$\beta$ emitters have lower luminosities than [O{\sc iii}] emitters, likely contributing more towards the faint-end of the luminosity function, but very little at the bright end. We discuss the implications in more detail in \S \ref{OIII_Hbeta_fluxes}.

\subsubsection{[O{\sc ii}] emitters at $z=2.2$}

We select [O{\sc ii}] emitters at $z=2.2$ with a similar selection as for H$\alpha$ emitters at $z=2.23$ in \citet{Sobral13}. Our selection is as follows: i) we disregard all candidate H$\alpha$ emitters and [O{\sc iii}]+H$\beta$ emitters obtained in the previous subsections (rejection of 3888 sources); ii) we use our photometric redshifts, applying the selection $1.8<z_{phot}<2.7$ (424 sources); and iii) we use a $BzK$ cut, to isolate $z>1.5$ galaxies \citep[][]{Sobral13}, adding 69 new sources to the sample. We also use spectroscopic redshift information to remove seven non-[O{\sc ii}] emitters in the sample and introduce two spectroscopically confirmed sources. We obtain a final sample of 488 [O{\sc ii}] emitters in SA22, with 19 sources being spectroscopically confirmed. Our spectroscopic sample is extremely limited for our [O{\sc ii}] sample, but based on the available redshifts, we estimate a $\sim90$\% completeness and a $<25$\% contamination. We note that we do not apply the Lyman-break selection in order to exclude $z>3$ emitters as in \citet{Sobral13}, because no major emission lines are found at slightly higher redshift --- but the reader is referred to \cite{Matthee14} for a full discussion of emitters at much higher redshift.


%
%
%
%
\begin{table}
\caption{Summary of the information in the SA22 catalog. It first shows the number of NB$_J$ detections, then the number of sources selected as emitters in Section 3.1. It shows the number of candidate emitters after removing spurious sources, the number of candidate stars and the final number of robust emitters. The remaining columns present the number of H$\alpha$, [O{\sc iii}]/H$\beta$ and [O{\sc ii}] candidates. When available, the number of spectroscopic redshifts are shown as well.}
\begin{center}
\begin{tabular}{lcc}
\hline
Sample & No. of sources & $z$-spec\\ \hline
NB$_J$ detections & 346,244 & 16964 \\  
Candidate emitters (after visual check) & 6315 & 541  \\   
Stars & 339 & 30 \\  
Robust emitters & 5976 & 511 \\   
\hline

H$\alpha$ ($z = 0.81$) & 2834 & 295 \\ 
{[O{\sc iii}]}+H$\beta$ ($z = 1.37/1.44$) & 1056  & 46 \\ 
{[O{\sc ii}]} ($z = 2.18$) & 488 & 19 \\ 
{[S{\sc ii}]} ($z=0.76$) & -- & 37 \\
$z<0.7$ ($z>3$ or unidentified) & 1638 & 82 (26) \\ 
\hline
\end{tabular}
\end{center}
\label{catalog}
\end{table}


\subsubsection{Final Samples: SA22}

Using the data and steps mentioned in previous chapters, we add an identifier in the range 0 to 3 to classify between other lines/unclassified (0), H$\alpha$ (1), [O{\sc iii}]+H$\beta$ (2) and [O{\sc ii}] (3) to the catalogue of emitters containing the final 5976 robust excess sources. Table \ref{catalog} presents the final numbers per sample, including the number of spectroscopically confirmed.

\subsubsection{Samples in COSMOS and UDS}

We supplement our sample using the HiZELS catalogue of line emitters available from \citet{Sobral13}, which has been derived in a very similar way to ours, and with almost identical data quality and depth. For H$\alpha$, we use the samples presented by \cite{Sobral13}: 425 H$\alpha$ emitters in COSMOS and 212 in UDS\footnote{By applying exactly the same selection as in SA22 we would obtain two extra sources in COSMOS, and three extra sources (and less six) in UDS relative to \cite{Sobral13}, but we chose to use the sample presented in \cite{Sobral13} for consistency. These minor differences in the sample make no difference in the results, particularly as they are found at the fainter end.} 

In order to select [O{\sc iii}]+H$\beta$ and [O{\sc ii}] emitters, we apply exactly the same criteria as in SA22. In COSMOS, we find 159 [O{\sc iii}]+H$\beta$ and 41 [O{\sc ii}] emitters. In UDS, we find 128 [O{\sc iii}]+H$\beta$ emitters at $z=1.4$ and 43 [O{\sc ii}] emitters at $z=2.2$. We note that the selection criteria (optimised for the COSMOS and UDS fields) in \cite{Khostovan15} to select [O{\sc iii}]+H$\beta$ and [O{\sc ii}] emitters with the HIZELS NB$_J$ sample is slightly different than ours in the photo-z selection (slightly more restrictive due to photo-zs in COSMOS and UDS being better than in SA22), but although it results in some very minor differences in the source numbers (maximum $\sim15$\% but typically within 5-10\%), the samples are essentially the same ($>90$\% of the samples are the same) and lead to the same results. Here we chose to use the same selection criteria as for SA22 to have fully consistent samples across fields, but we note that even if we applied \cite{Khostovan15} selection criteria our results would not change.

We also note that the HiZELS data \citep{S09a,Sobral13} are slightly deeper (by $\sim0.1$\,dex in luminosity) than our SA22 survey and that the narrow-band filter used by \cite{Sobral13} is slightly wider, and thus naturally recovers a higher number of emitters per deg$^2$ if that is not taken into account.

We provide a summary of the sample in the three fields in Table \ref{catalog_samples}. The final samples (SA22, COSMOS and UDS) are by far the largest ever assembled, yielding 3471 H$\alpha$ emitters at $z\sim0.8$ ($\sim400$ spectroscopically confirmed),1343 [O{\sc iii}]+H$\beta$ at $z\sim1.4$ and 572 [O{\sc ii}] emitters at $z\sim2.2$. Some relatively significant variations in source densities are found across fields and within fields. This is evaluated in \S5, where we present how important is cosmic variance on different scales for each one of the emission-lines.

%

%
%
%
%
%
\begin{table}
\caption{A summary of our final samples of H$\alpha$, {[O{\sc iii}]}+H$\beta$ and {[O{\sc ii}]} emitters at $z\sim0.8$, $z\sim 1.4$ and $z \sim 2.2$, respectively. }
\begin{center}
\begin{tabular}{cccc}
\hline
Samples & No. of sources & Volume  & Depth (log$_{10}$L) \\
H$\alpha$ ($z\sim0.8$) & (\#) & ($10^{6}$\,Mpc$^3$) & (erg\,s$^{-1}$) \\  
 \hline
SA22 & 2834 & 0.9 & 41.35 \\  
COSMOS & 425 & 0.1 & 41.25 \\  
UDS & 212 & 0.08 & 41.30 \\  
Full Sample & 3471 & 1.1 & 41.35 \\  
\hline
{[O{\sc iii}]}+H$\beta$ ($z\sim 1.4$) & (\#) & ($10^{6}$\,Mpc$^3$) & (erg\,s$^{-1}$) \\  
 \hline
SA22 & 1056 & 2.48 & 42.0 \\ 
COSMOS & 159 & 0.33 & 41.90 \\ 
UDS & 128 & 0.24 & 41.95 \\ 
Full Sample & 1343 & 3.05 & 42.0 \\   
\hline
{[O{\sc ii}]} ($z \sim 2.2$) & (\#) & ($10^{6}$\,Mpc$^3$) & (erg\,s$^{-1}$) \\  
 \hline
SA22 & 488 & 2.62 & 42.6 \\
COSMOS & 41 & 0.29 & 42.50 \\
UDS & 43 & 0.22 & 42.55 \\
Full Sample & 572 & 3.12 & 42.6\\

\hline
\end{tabular}
\end{center}
\label{catalog_samples}
\end{table}

\subsection{AGN fraction for line emitters} \label{AGN}

We take advantage of the similar samples we selected in COSMOS and UDS to estimate the fraction of potential AGNs in our SA22 sample (where no {\it Chandra} data and no IRAC data are available). We start by using C-COSMOS \citep{Elvis09} to find that, in agreement with \cite{Garn2010a}, only $\sim1$\% of H$\alpha$ sources are detected in the X-rays (5 sources, X-ray luminosities of $10^{42.8\pm0.11}$\,erg\,s$^{-1}$). For our [O{\sc iii}]+H$\beta$ and [O{\sc ii}] samples in COSMOS, we only find one X-ray match per sample (both sources with X-ray luminosities of $\sim10^{43.4}$\,erg\,s$^{-1}$), thus resulting in a very small fraction ($<1$\,\% for [O{\sc iii}]+H$\beta$ at $z\sim1.4$) and $\sim2-3$\% for [O{\sc ii}] at $z=2.23$. It is therefore clear that the fraction of X-ray detected AGNs in our samples are only at the level of $\sim1$\,\%. All X-ray AGN within our sample are found to have intermediate line luminosities (so they are not the highest luminosity line emitters, but none is a weak line emitter).

We also use deep IRAC data, in both COSMOS and UDS, to look for sources dominated by signatures of the stellar bump (star-forming dominated), thus showing a blue colour beyond $\sim1.6$\,$\mu$m rest-frame and sources dominated by signatures of a red power-law (AGN dominated), showing red colours even beyond $\sim1.6$\,$\mu$m rest-frame \citep[see e.g.][]{Garn2010a}. Given the typical errors on IRAC photometry, particularly for redder bands, and motivated by the colours of our X-ray AGN (in order to recover the majority of them with the appropriate IRAC colours given the redshift of the sources), we apply the following cuts to select potential AGN, which should be mostly interpreted as an upper limit to the AGN contamination. For H$\alpha$ ($z=0.8$): [3.6]-[4.5]$>0.0$ (10\% potential AGN); for [O{\sc iii}]+H$\beta$ ($z\sim1.4$): [4.5]-[5.8]$>0.1$ (18\% potential AGN) and for [O{\sc ii}] ($z=2.2$): [5.8]-[8.0]$>0.2$ (23\% potential AGN). The contamination for these relatively ``typical'' sources can be well approximated by a constant contamination across luminosities at least up for the luminosities probed by the UDS and COSMOS samples. 

In order to constrain the fraction of AGN among the most luminous H$\alpha$ emitters, we use the results from \cite{Sobral15}. \cite{Sobral15} present the results of the spectroscopic follow-up of very luminous H$\alpha$ emitters, allowing to constrain the fraction of AGN for H$\alpha$ luminosities $L>L^*$ (up to $L\sim50L^*$). Their results show that while the AGN fraction is relatively constant up to $L\sim L^*$ (around 10\%), it strongly correlates with $L/L^*$ for higher luminosities, reaching $\sim100$\% by $L\sim50L^*$. Therefore, for H$\alpha$ emitters, we use results from X-rays and from IRAC, which allow us to have a global AGN fraction for typical luminosities, and the results from \cite{Sobral15}, which allow us to estimate the AGN fraction at the highest luminosities (for H$\alpha$ emitters). We thus assume that our H$\alpha$ sample ($z=0.8$) will likely be $\sim10$\% contaminated by AGN up to $L\sim L^*$, and we use the AGN fraction as a function of $L/L^*$ found by \cite{Sobral15}, which is given by $0.59\times\log_{10}(L/L^*)+0.112$.

For our sample of [O{\sc iii}]+H$\beta$ and [O{\sc ii}] emitters we have no information regarding the AGN fraction of the highest luminosity emitters. We thus assume a constant AGN fraction when converting luminosity densities to star formation rate densities. For our sample of [O{\sc iii}]+H$\beta$ we assume we will be $\sim15$\,\% contaminated by AGNs \citep[][find that for H$\alpha$ emitters at $z=1.4$ only 10\% are AGN]{Stott13b} and for our [O{\sc ii}] sample at $z=2.2$ we use a correction of $\sim20$\,\%.

%
%
%
%
\begin{figure}
\centering
\includegraphics[width=8cm]{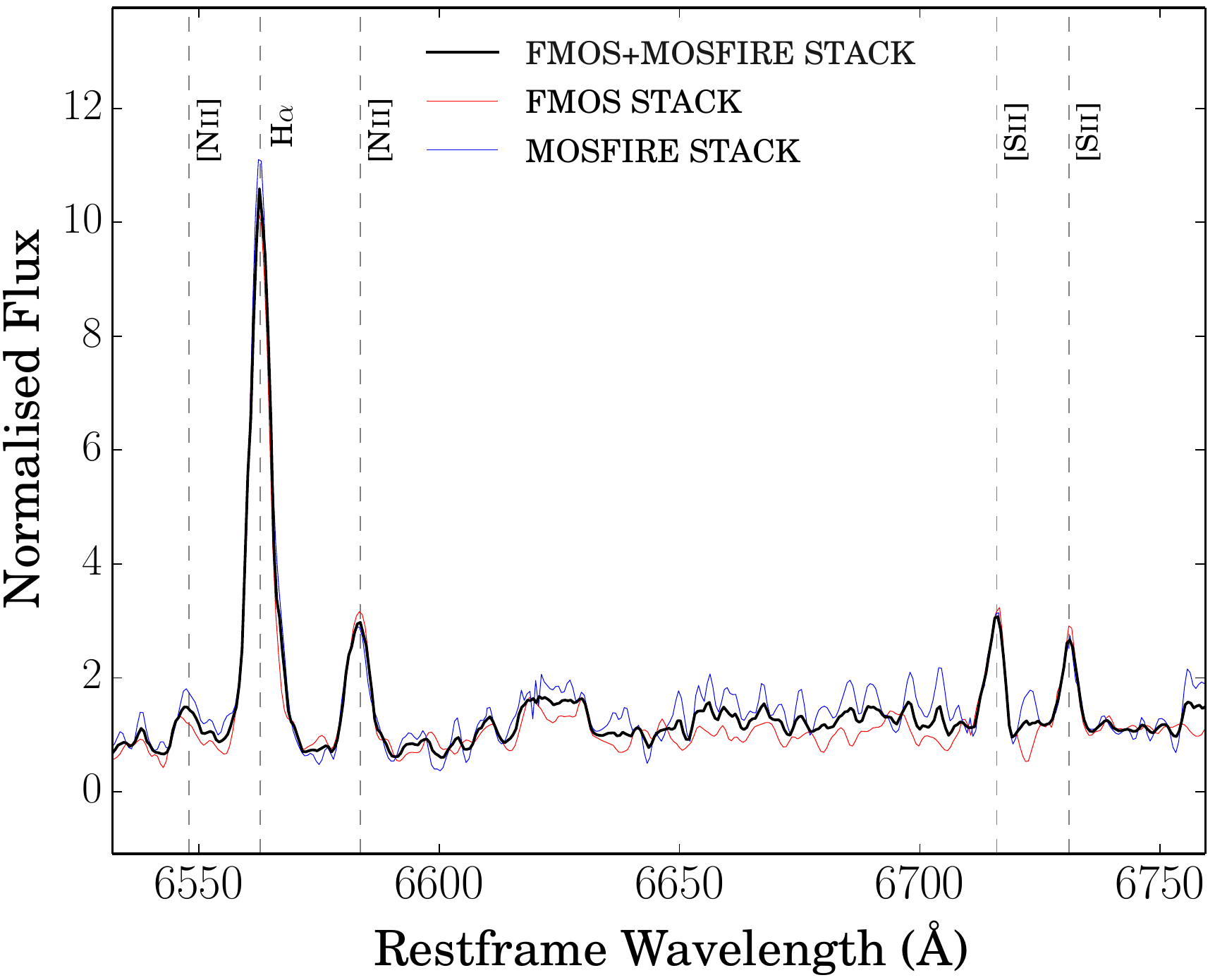}
\caption{Stacked spectra of $z=0.81$ H$\alpha$ emitters obtained with FMOS and MOSFIRE. As a whole, our H$\alpha$ emitters have a metallicity $\rm 12+log_{10}(O/H)=8.56\pm0.05$, slightly sub-solar (but consistent with solar metallicity $12+\rm log_{10}(O/H)=8.66\pm0.05$). We also make significant detections of the [S{\sc ii}] doublet, with a line ratio of I([S{\sc ii}]$_{6716}$)/I([S{\sc ii}]$_{6731}) = 1.33 \pm 0.08$, implying an electron density of  40-200\,cm$^{-3}$. The [S{\sc ii}]$_{6716}$/I(H$\alpha$)$=0.14\pm0.02$ ratio also implies ionisation potential of log$_{10}({\rm U})=-3.9\pm0.5$\,cm$^{-3}$.}
\label{Stack_spectra}
\end{figure}

%
%
%
%
%

%
%
%
%
%
%
%

\section{METHODS: COMPLETENESS and CORRECTIONS}

%
%
%
\begin{figure}
\centering
\includegraphics[width=8cm]{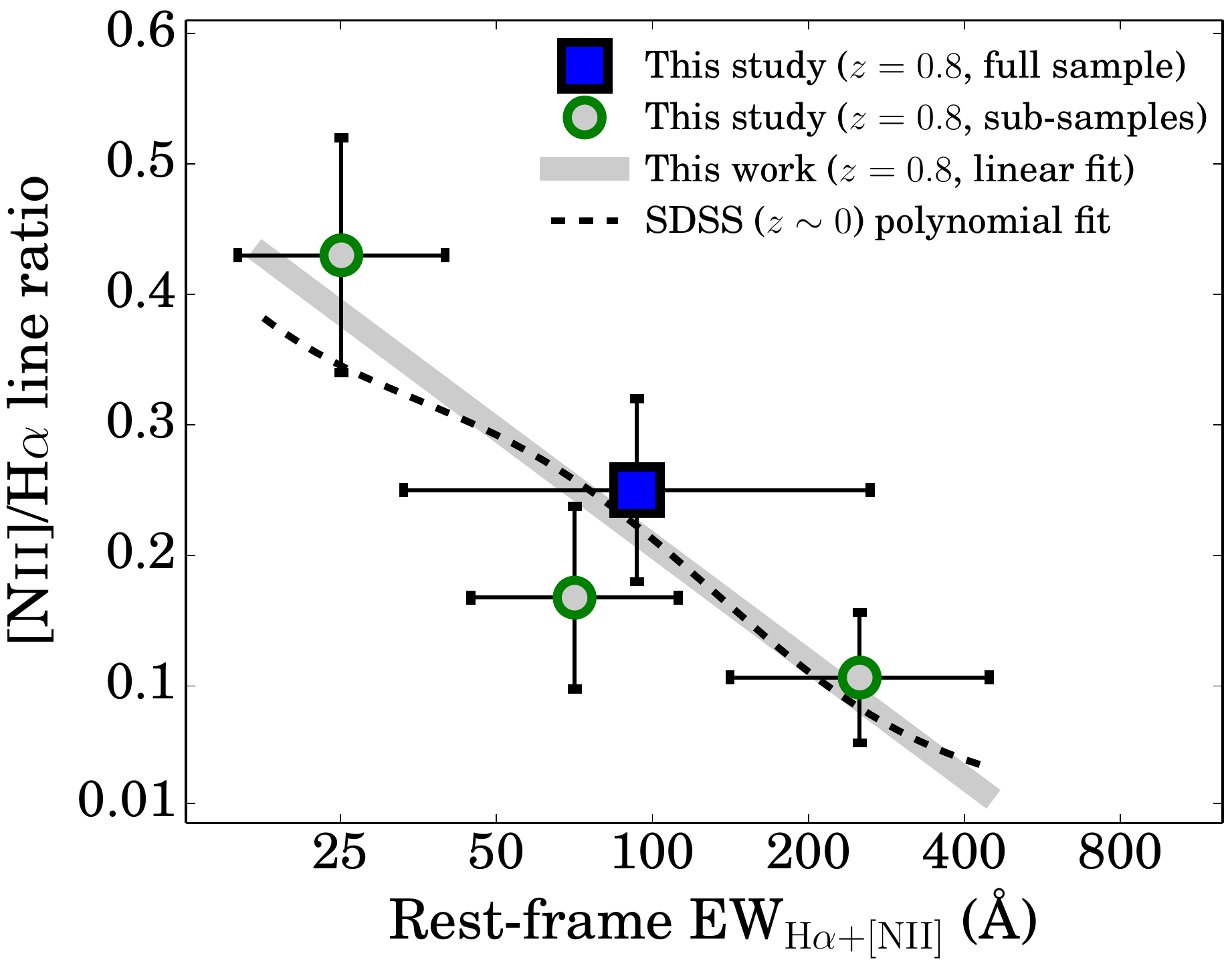}
\caption{\small{The observed anti-correlation between the [N{\sc ii}]6583\AA/H$\alpha$ line ratio and the rest-frame EW(H$\alpha$+[N{\sc ii}]6583\AA) at $z=0.81$ in this study and a comparison with the polynomial fit to SDSS at low redshift. Given the range of EWs and line ratios probed, we find that while the SDSS fit is a good fit to the data at $z=0.81$, a linear fit does equally well, and is a much simpler way to robustly estimate [NII]/H$\alpha$. The relation is $f(\rm [NII]/H\alpha)=-0.296\times\log_{10}(\rm EW_{\rm H\alpha+[NII]})+0.8$. We show three bins which split the sample in 3 relatively evenly in parameter space and that do not overlap, together with the stack for the full sample, and 3 other stacks which split the sample in terms of number of H$\alpha$ emitters.}}
\label{NII_Ha_EW}
\end{figure}

\subsection{Flux corrections: [NII]/H$\alpha$ correction} \label{EW_CorrectionNII}

When computing line fluxes and equivalent widths for the H$\alpha$ emitters (H$\alpha$ line), one must note that the adjacent [N{\sc ii}] lines at 6548\AA\ and 6583\AA\ will also contribute to both quantities, increasing them both (see e.g. Figure \ref{Stack_spectra}). [N{\sc ii}]6583\AA\ is clearly the strongest and the most important to take into account (see Figure \ref{Stack_spectra}); we refer to it as simply [N{\sc ii}] for the remaining of the paper. One way to correct for this is to use the SDSS relation between F$_{{[\rm NII]}}/$F$_{{\rm H}\alpha}$ and the total measured equivalent width EW(H$\alpha$+[N{\sc ii}]), \citep[][]{Villar08,Sobral12}. Given the representative spectroscopic follow-up of our H$\alpha$ emitters, we can directly test whether the polynomial correction presented in \citet{Sobral12} (based on SDSS) is appropriate for $z\sim1$.

We use our FMOS and MOSFIRE data and stack as a function of rest-frame EW(H$\alpha$+[N{\sc ii}]) (measured from the NB data as that is what we want to evaluate). We recover a clear anti-correlation between F$_{{[\rm N\sc II]}}/$F$_{{\rm H}\alpha}$ and rest-frame EW(H$\alpha$+[N{\sc ii}]), as shown in Figure \ref{NII_Ha_EW}. We find that the trend is fully consistent with SDSS of decreasing F$_{{[\rm N\sc II]}}/$F$_{{\rm H}\alpha}$ as a function of EW(H$\alpha$+[N{\sc ii}]). We note, as shown in Figure \ref{NII_Ha_EW}, that a linear relation is an even simpler correction which is extremely similar to the SDSS relation. The linear relation that we derive is: 
\begin{equation}
f(\rm [NII]/H\alpha)=-0.296\times\log_{10}(\rm EW_{\rm H\alpha+[NII]})+0.8
\end{equation}
and is valid for rest-frame EW(H$\alpha$+[N{\sc ii}]) from $\sim15$\,\AA \ to $\sim600$\,\AA \ (see Figure \ref{NII_Ha_EW}). 

We note that if a single correction is applied (for similar surveys), then the value to be applied for F$_{{[\rm N\sc II]}}/$F$_{{\rm H}\alpha}$ should be 25\%, as the F$_{{[\rm N\sc II]}}/$F$_{{\rm H}\alpha}$ for the full stack (median stack, see Figure \ref{Stack_spectra}) is $0.25\pm0.05$. This suggests \citep[in agreement with e.g.][]{Swinbank12,Stott13b,Sobral13b}, using the \cite{PettiniPagel04} calibration, that as a whole, our H$\alpha$ emitters have a metallicity $12+\rm log_{10}(O/H)=8.56\pm0.05$, slightly sub-solar (solar metallicity $12+\rm log_{10}(O/H)=8.66\pm0.05$).

In our stacked spectra, we also make significant detections of the [S{\sc ii}] doublet. We find a median line ratio of I([S{\sc ii}]$_{6716}$)/I([S{\sc ii}]$_{6731}) = 1.33 \pm 0.08$, which implies an electron density of 40-200\,cm$^{-3}$ \cite[][]{Osterbrock89}. We also find that [S{\sc ii}]$_{6716}$/I(H$\alpha$)$=0.14\pm0.02$, which implies an ionisation potential of log$_{10}({\rm U})=-3.9\pm0.5$\,cm$^{-3}$ \cite[][]{Osterbrock89}. However, we emphasise that the stacked spectra provide only information on the median value, and no indication on the range of values.

\subsection{The relative contributions from [O{\sc iii}]$_{5007}$, [O{\sc iii}]$_{4959}$ and H$\beta$ to the sample of [O{\sc iii}]+H$\beta$ emitters at $z\sim1.4$} \label{OIII_Hbeta_fluxes}

For [O{\sc iii}]+H$\beta$, we start by noting that based on our spectroscopic follow up, $\sim16$\% of the [O{\sc iii}]+H$\beta$ emitters turned out to be H$\beta$ and the rest to be [O{\sc iii}]. We also find that spectroscopically confirmed H$\beta$ emitters are found to have lower luminosities than spectroscopically confirmed [O{\sc iii}] emitters. This, combined with the low fraction within the sample, would be a relatively good motivation towards ignoring such emitters \citep[this is done by most studies, e.g.][]{Ly2007}. However, attempting to further split the H$\beta$ and [O{\sc iii}] luminosity functions, we take the following approach. We use the \cite{Sobral13} $z=1.47$ H$\alpha$ luminosity function, and, with a simple assumption of A$_{\rm H\alpha}=1$\,mag and case B recombination, we predict the H$\beta$ Luminosity function at $z\sim1.44$ (also taking into account what flux we would recover with our NB filter). We find that the number densities of H$\beta$ emitters are significantly below the number densities of [O{\sc iii}]+H$\beta$ emitters. Given our depth, this simple prediction is very much in line with the 16\% fraction of H$\beta$ found (and with H$\beta$ emitters being preferentially found at fainter luminosities).

%
%
%
%
\begin{figure}
\centering
\includegraphics[width=8.2cm]{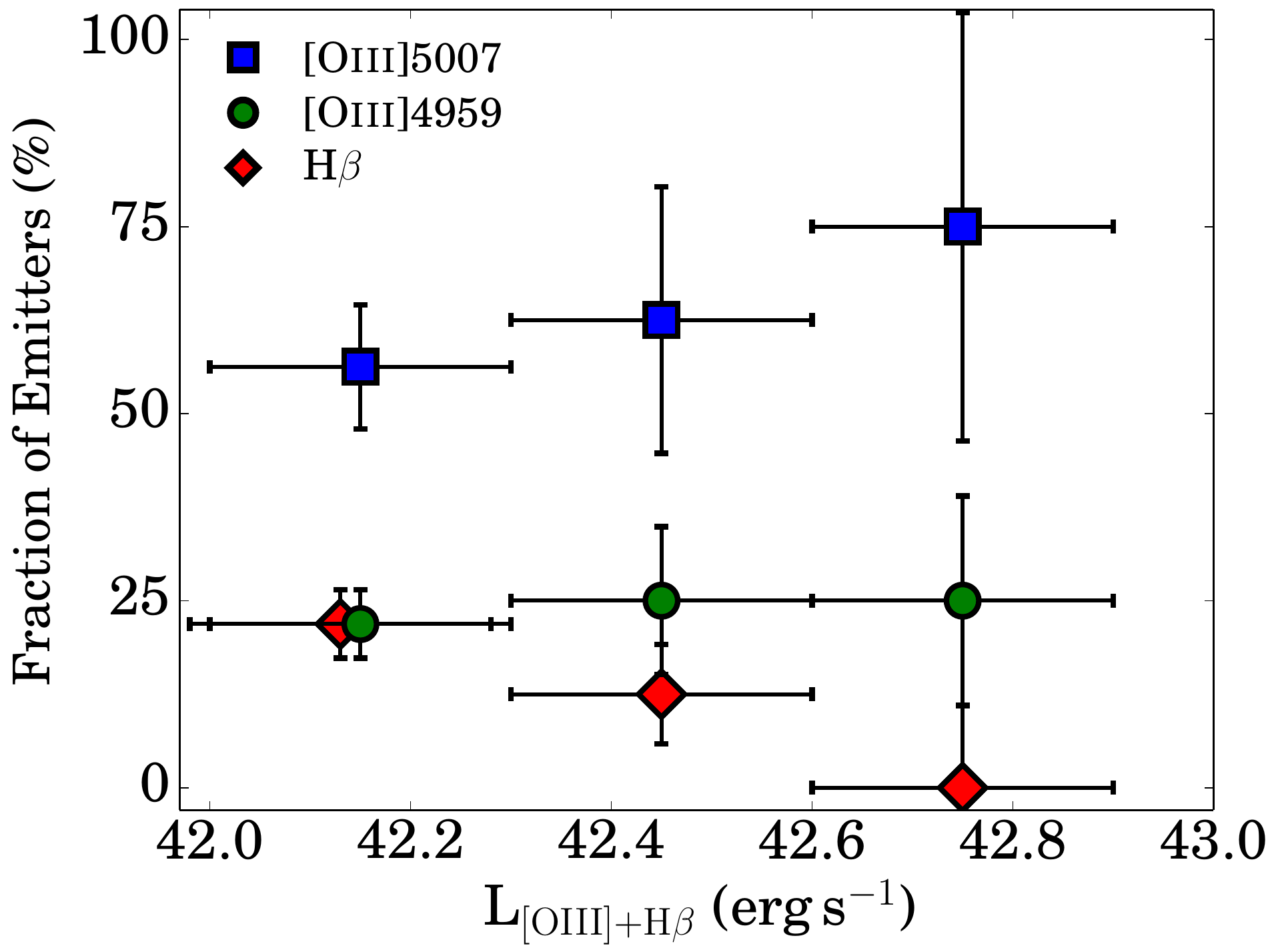}
\caption{The fraction (within the spectroscopic sample) of [O{\sc iii}]5007, [O{\sc iii}]4959 and H$\beta$ emitters as a function of [O{\sc iii}]+H$\beta$ luminosity within the [O{\sc iii}]+H$\beta$ sample of emitters at $z\sim1.4$. We find that [O{\sc iii}]5007 emitters dominate at all luminosities, and that H$\beta$ emitters only start to have a small contribution for lower luminosities.} 
\label{OIII_Hb_fractions}
\end{figure}

Due to the separation in wavelength, it is not possible for both H$\beta$ and [O{\sc iii}] to contribute to the NB flux measured. We note, nonetheless, that there is a narrow range of redshifts where both lines from the [O{\sc iii}] doublet are detected (see one example of an emitter in Figure \ref{EXAMPLES} where both [O{\sc iii}] lines contribute to the narrow-band flux) at opposite wings of the filter. Indeed, the [O{\sc iii}]${4959}$ emission line is found to be even more important within the sample than H$\beta$ and with important consequences for deriving luminosity functions. While we find that [O{\sc iii}]${5007}$ (only) dominates the sample of spectroscopically confirmed [O{\sc iii}]+H$\beta$ emitters, representing $\sim50$\% of the full sample, [O{\sc iii}]${4959}$ (only) represents 27\% (see e.g. Figure \ref{EXAMPLES}), almost twice as common as H$\beta$ and 7\% of the sources are actually detections of both [O{\sc iii}] lines at opposite wings of the filter (e.g. Figure \ref{EXAMPLES}). Within the spectroscopic sample of [O{\sc iii}]+H$\beta$ emitters, we study the fraction of each emitter as a function of luminosity and show the results in Figure \ref{OIII_Hb_fractions}. Our analysis of the spectroscopically confirmed emitters shows that [O{\sc iii}]${5007}$ emitters are distributed in a very Schechter like form, at all luminosities, H$\beta$ are only found at lower luminosities, but [O{\sc iii}]${4959}$ emitters, and simultaneous detections of [O{\sc iii}]${5007}$ and [O{\sc iii}]${4959}$ lines, are found both at the faintest luminosities (in the NB) and at the brightest luminosities, thus significantly boosting number counts at both the faintest luminosities (likely steepening the faint-end slope) and at the brightest luminosities (likely contributing to a non-Schechter form).

Our results are very important in order to interpret the results from the luminosity function in \S \ref{OIII_Hbeta} and for other similar studies \citep[e.g.][]{Ly2007,Khostovan15}. Thus, while we present a [O{\sc iii}]+H$\beta$ luminosity function, we caution that in order to properly derive luminosity functions for each of the individual lines contributing to it, a much more detailed spectroscopic follow-up is needed, and that simply assuming that similar samples will be dominated by [O{\sc iii}]${5007}$ (ignoring, for example, the [O{\sc iii}]${4959}$ line, or simultaneous measurements of the doublet), will lead to strong systematic errors/biases. We therefore refer to these emitters as [O{\sc iii}]+H$\beta$ throughout the paper. We also take our results into account when estimating the total volume probed by our survey for [O{\sc iii}]+H$\beta$ emitters. Given that [O{\sc iii}]${5007}$ dominates the sample, we use the full volume probed with that emission line, but we correct for the expected addition of extra emitters (picked up over extra volumes). Thus, we also add to the total volume probed (because we are also sensitive to the other lines, and following our spectroscopic results) 16\% of full H$\beta$ volume probed by us and 25\% of the volume probed for [O{\sc iii}]${4959}$ emitters.

\subsection{Luminosity Calculations and Extinction corrections} \label{lfalpha}

In order to calculate luminosity functions for our samples, line fluxes are converted to luminosities by applying:

\begin{equation}
  L_{{\rm line}}=4\pi {\rm D_L^2}{\rm F}_{{\rm line}}
\end{equation}
where ${\rm D_L}$ is the luminosity distance. We use ${\rm D_L}$=5367\,Mpc for H$\alpha$ emitters at $z=0.81$, ${\rm D_L}$=9752.7\,Mpc for [O{\sc iii}]/H$\beta$ emitters at $z=1.4$ and ${\rm D_L}$=17746.5\,Mpc for [O{\sc ii}] emitters at $z=2.2$.

We note that all our luminosity functions are observed luminosity functions, not dust corrected. We only apply extinction corrections when converting luminosity density to star formation rate density. We also note that for the [O{\sc ii}] sample, we do not need to apply or investigate any correction, as no other line (close enough in rest-frame wavelength) is expected to contribute to the flux measured.

\subsection{Completeness corrections}
\label{compl}

Fainter sources and those with weak emission lines might be missed and thus not included in the sample; this will result in the underestimation of the number of emitters, especially at lower luminosities. In order to account for that we follow \citet{Sobral13} to estimate completeness corrections per sub-field per emission line. Very briefly, we use sources which have not been selected as line emitters ($\Sigma<3$ or EW$<30$\, {\AA}) and that have a photo-$z$ within $\pm0.5$ of the appropriate line of interest (thus, we use different samples to estimate the completeness of different lines; see e.g. \citet{Sobral12}). We then add emission-line flux to all those sources, and study the recovery fraction as a function of input flux. We do these simulations in a sub-field by sub-field basis. We then apply those corrections in order to obtain our completeness-corrected luminosity functions.

\subsection{Filter profile corrections} \label{profil_filt}

The narrow-band filter transmission function is not a perfect top-hat, so the real volume surveyed is a function of intrinsic luminosity. For example, luminous line emitters will be detectable over a larger volume than the fainter ones, as they can be detected in the wings of the filters (although they will be detected as fainter sources in these cases). Low luminosity sources, however, will only be detectable in the central regions of the filter, leading to a smaller effective volume.

In order to correct for this when deriving the luminosity functions, we follow \citet{Sobral12}. Firstly, we compute the luminosity function assuming a top-hat narrow-band filter. We then generate a set of $10^{10}$ line emitters with a flux distribution given by the measured luminosity function, but spread evenly over the redshift range being studied (assuming no cosmic structure variation or evolution of the luminosity function over this narrow redshift range). We fold the fake line emitters through the top-hat filter model to confirm that we recover the input luminosity function perfectly. Next, we fold the fake line emitters through the real narrow-band profiles -- their measured flux is not only a function of their real flux, but also of the transmission of the narrow-band filter for their redshift. The simulations show that the number of brighter sources is underestimated relative to the fainter sources. A mean correction factor between the input luminosity function and the one recovered (as a function of luminosity) was then used to correct each bin.

\section{RESULTS}

\subsection{Luminosity functions} \label{lf}

The estimate of the source density in a luminosity bin of width $\Delta(\log L)$ centred on $\log L_c$ is given by the sum of the inverse volumes of all the sources in that bin, after correcting for completeness. The volume probed is calculated taking into account the survey area and the narrow-band filter width, followed by applying the appropriate real filter profile corrections obtained in \S\ref{profil_filt}.

The luminosity functions presented here are fitted with Schechter functions defined by the three parameters: $\alpha$, $ \phi ^*$ and $L^*$ (see Table \ref{SFRD_LFS}). For the H$\alpha$ luminosity function, we can still get a reasonable constraint on $\alpha$, but the data are too shallow to do the same for [O{\sc iii}]+H$\beta$ and [O{\sc ii}], and thus for those we concentrate on just fitting for $ \phi^*$ and L$^*$ (and fixing $\alpha$ to the values commonly used in the literature for comparison). The final luminosity functions are presented in Figures~\ref{HalphaLF} and \ref{OIII_OII_LF}, and in Tables~\ref{LF_HA_TAB}, \ref{LF_OIII_TAB} and \ref{LF_OII_TAB}. We note that we do not apply any dust correction prior to fitting the luminosity functions -- dust corrections are only applied when converting luminosity densities to estimates of the star formation rate density. The best fit Schechter functions are presented in Table \ref{SFRD_LFS}.

\subsubsection{H$\alpha$ luminosity function at $z=0.81$}

\begin{figure*}
\centering
\includegraphics[width=15cm]{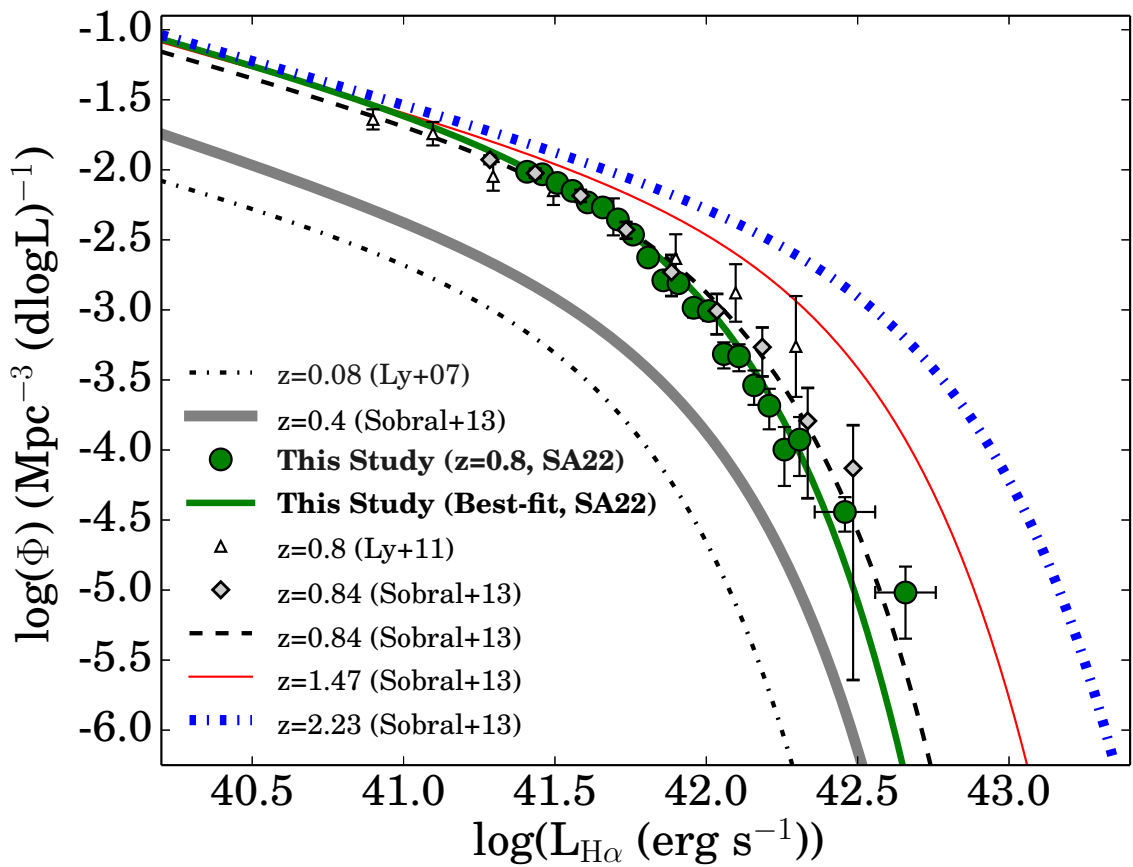}
\caption{Observed H$\alpha$ luminosity function at $z=0.81$ from our survey, comparison with previous surveys at a similar redshift, and evolution of the luminosity function (none of the luminosity functions have been corrected for dust extinction). We find a very good agreement between our results and those in the literature (well within the errors excepted from cosmic variance), but are able to extend the H$\alpha$ luminosity function to much higher luminosities, as well as reducing the Poissonian errors significantly. The errors due to cosmic variance are $<10$\% at most. Our results agree with the strong L$^*$ evolution with increasing redshift.} 
\label{HalphaLF}
\end{figure*}

We present our H$\alpha$ luminosity function at $z=0.81$ in Figure~\ref{HalphaLF} and Table~\ref{LF_HA_TAB}. We confirm a strong evolution in the H$\alpha$ luminosity function from $z=0$ to $z=0.8$. We also find good agreement with both \cite{Sobral13} and \cite{CHU11}, as can be seen by directly comparing the data points and the best fit luminosity function from e.g. \cite{Sobral13}. However, as Figure \ref{HalphaLF} clearly shows, we are able to not only extend the LF to higher luminosities, and to constrain the number densities of the most luminous H$\alpha$ emitters with unprecedented accuracy, but also determine it with 3$\times$ more bins while reducing the Poissonian errors significantly at every bin. While the agreement is relatively good at all luminosities, we find a slightly lower L$^*$ or slightly lower $ \phi^*$, in line with e.g. \cite{Colbert}. Our results in \S5.3 also show that we over-come cosmic variance, with the errors becoming a few times smaller than all previous surveys, and that the differences with respect to \cite{Sobral13} and \cite{CHU11} are fully explained by cosmic variance. Our results also confirm that the faint-end slope is relatively steep (we find $\alpha=-1.6\pm0.2$), in excellent agreement with both \cite{Sobral13} and \cite{CHU11} and that L$^*$ evolves significantly from $z\sim0$ to $z\sim0.8$.
We also note that our spectroscopically confirmed sub-sample (301 spectroscopically confirmed H$\alpha$ emitters in the SA22 sample) fully supports our results. Moreover, we find that the density of the most luminous emitters is still consistent with \cite{Sobral13}, although, as shown in \cite{Sobral15}, the most luminous H$\alpha$ emitters at $z\sim0.8$ have a significant fraction of AGN, particularly broad-line AGN.

%
%
%
%

\subsubsection{[OIII]+H$\beta$ luminosity function at $z\sim1.4$} \label{OIII_Hbeta}

\begin{figure*}
\centering
\begin{tabular}{ccc}
\includegraphics[width=8.5cm]{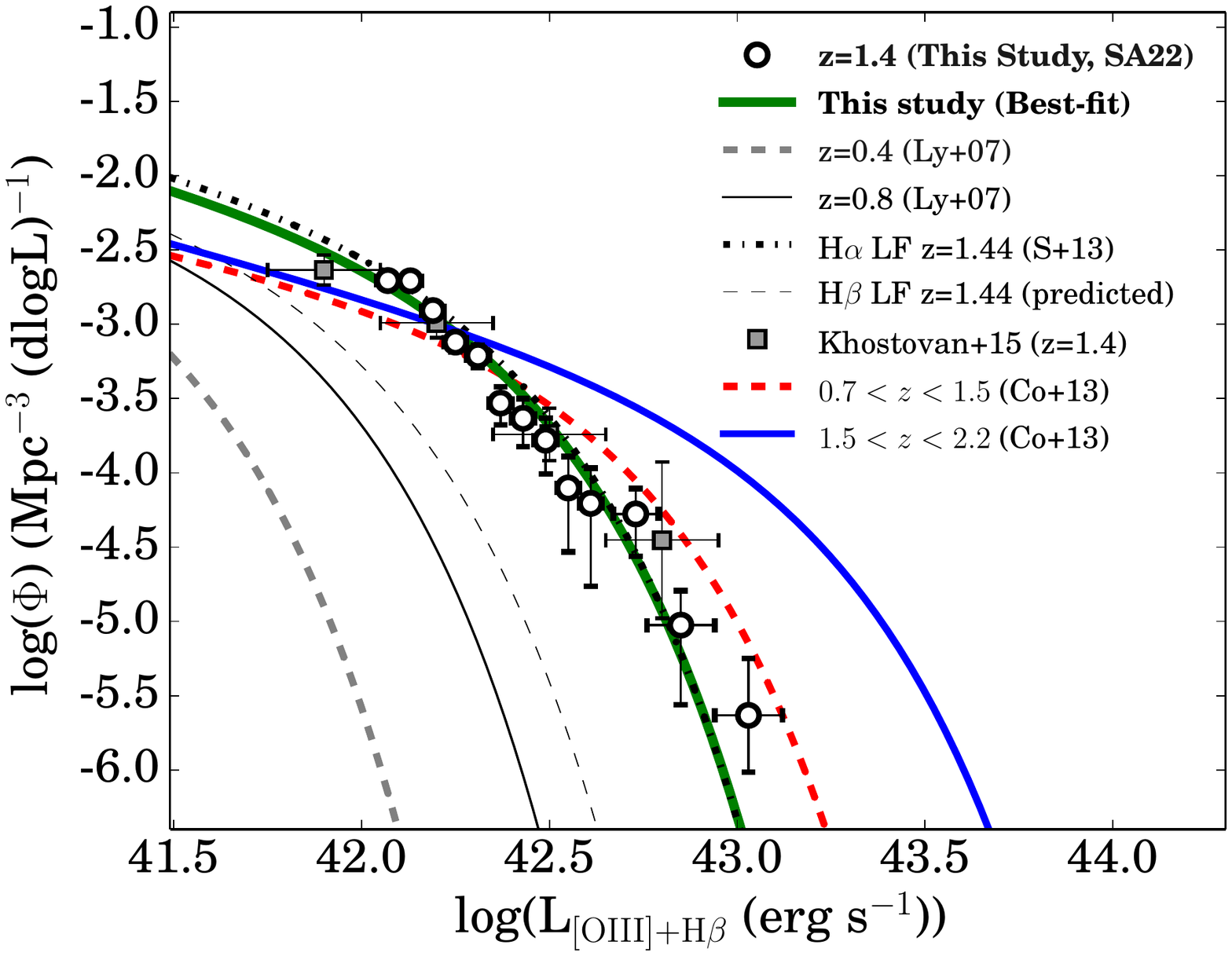}&
\includegraphics[width=8.5cm]{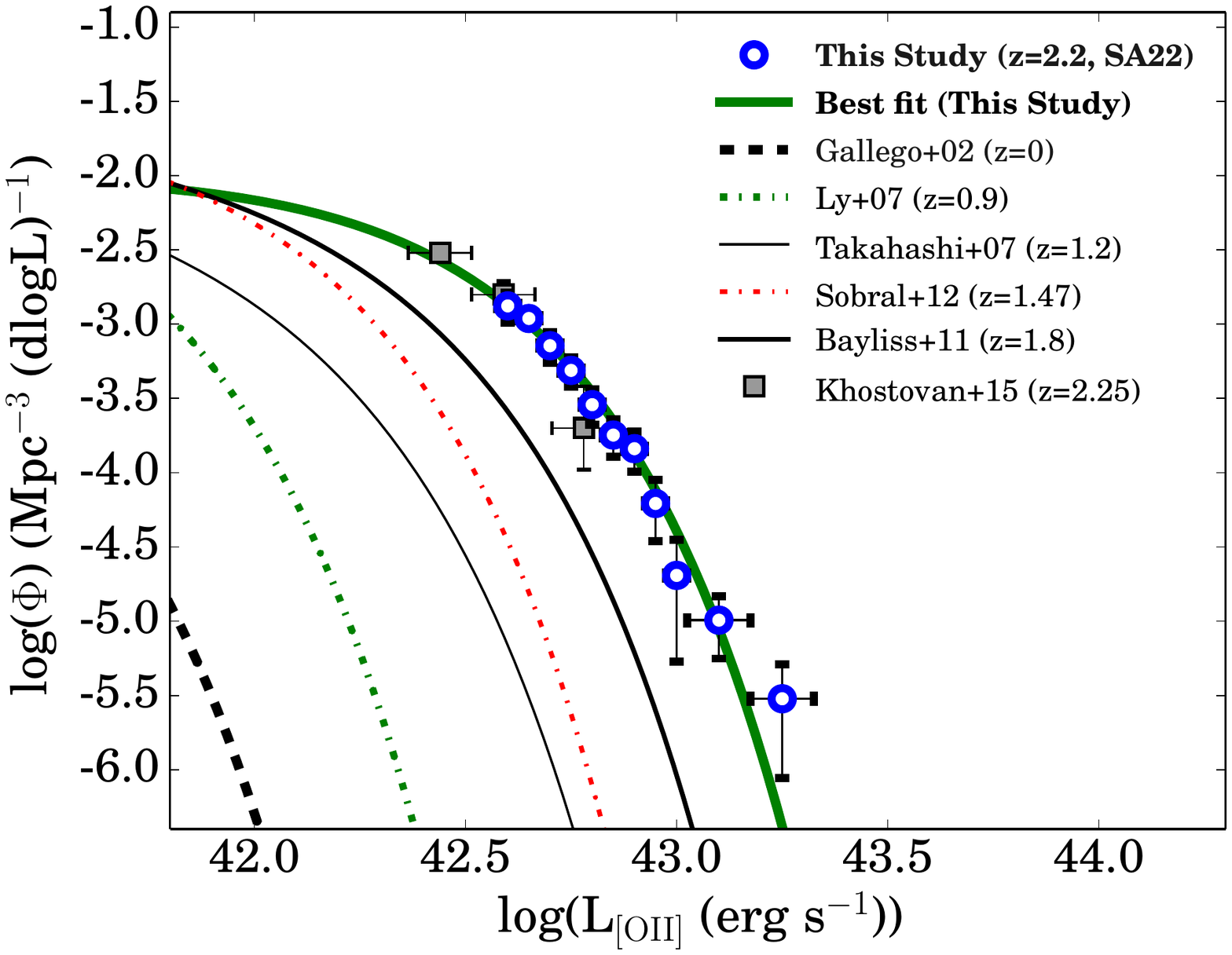}\\
\end{tabular}
\caption{{\it Left:} Observed [O{\sc iii}]+H$\beta$ luminosity function at $z\sim1.4$ from our SA22 survey and comparison with other studies/surveys (including the LF from UDS+COSMOS derived by \citealt{Khostovan15}). Our results clearly show that the [O{\sc iii}]+H$\beta$ luminosity function is evolving very strongly with redshift, mostly due to a strong increase in L$^*$, similarly to the behaviour of the H$\alpha$ luminosity function. Our best-fit is very similar to the observed H$\alpha$ luminosity function at the same redshift (also shown), and, at the faint end, reveals a higher number of [O{\sc iii}]+H$\beta$ emitters than in \citet{Colbert} -- these are likely explained by a combination of H$\beta$ emitters (we also show our predicted H$\beta$ luminosity function based on the H$\alpha$ luminosity function from \citet{Sobral13}), but also due to the prevalence of the [O{\sc iii}]${4959}$ line (see Section \ref{OIII_Hbeta_fluxes}). The deviations from a Schechter function are likely explained by the mix of different lines contributing to the final LF, as the observed luminosity function is likely a combination of three different functions (for the three different emission lines). For higher luminosities, our data is in good agreement with the results from \citet{Colbert} and our slightly lower number densities are likely a result of accounting from the contribution of both [O{\sc iii}]4959 and H$\beta$ for the final volue probed. Most importantly, we compare our results with \citet{Khostovan15}, who present the [O{\sc iii}]+H$\beta$ LF from HiZELS in COSMOS+UDS, and find a very good agreement. {\it Right:} [O{\sc ii}] luminosity function at $z=2.2$ from our SA22 survey and comparison with other lower redshift surveys such as \citet{GalegoOII}, \citet{Takahashi07} and \citet{Bayliss}. Similarly to the evolution of the H$\alpha$ and [O{\sc iii}]+H$\beta$ luminosity functions, the [O{\sc ii}] luminosity function shows a strong L$^*$ evolution. Such increase in L$^*$ with increasing redshift is even stronger by a few times than what is seen for other lines. On the other hand, there is little to no evidence of evolution in $ \phi^*$. We also compare our results from SA22 with those from UDS and COSMOS (HiZELS) presented by \citet{Khostovan15} and find a very good agreement, at both the bright and faint ends.} 
\label{OIII_OII_LF}
\end{figure*}

We present the observed (non-dust corrected) [O{\sc iii}]+H$\beta$ luminosity function at $z\sim1.4$ in Figure \ref{OIII_OII_LF} and in Table \ref{LF_OIII_TAB}. We have fixed $\alpha=-1.6$ for the best-fit. We find a clear luminosity evolution from $z\sim0$ to $z\sim1.4$ when compared to other luminosity functions from the literature, in line with what is seen for H$\alpha$. Interestingly, if one uses empirical conversions from [O{\sc iii}] to H$\alpha$ \citep[e.g.][]{Ly2007}, [O{\sc iii}]/H$\alpha\approx1.05$, and plot the H$\alpha$ LF at $z\sim1.4$, we find that it resembles the [O{\sc iii}]+H$\beta$ luminosity function surprisingly well. In other words, [O{\sc iii}]+H$\beta$ emitters at $z\sim1.4$ have similar number densities to H$\alpha$ emitters at the same redshift. However, as noted in \S \ref{OIII_Hbeta_fluxes}, the [O{\sc iii}]+H$\beta$ luminosity function is far from being simple/easy to interpret (as it is a composite of a few different lines, unlike H$\alpha$).

Comparing with the largest {\it HST} slitless grism survey \citep{Colbert}, we find very good agreement at moderate to high luminosities, but we recover more line emitters at the faintest luminosities, as discussed earlier in \S \ref{OIII_Hbeta_fluxes}. Based on our spectroscopic data, our [O{\sc iii}]+H$\beta$ luminosity function is likely a sum of three different luminosity functions, all contributing in varying forms across the observed luminosities, thus showing a non-Schechter like form: [O{\sc iii}]${4959}$, [O{\sc iii}]${5007}$, [O{\sc iii}]${4959}$+[O{\sc iii}]${5007}$ (where both lines are both being detected and measured by the narrow-band filter) and H$\beta$.

%
%
%
%
%
%

\subsubsection{[OII] luminosity function at $z=2.2$}

We present the observed (non-dust corrected) [O{\sc ii}] luminosity function at $z=2.2$ in Figure \ref{OIII_OII_LF} and in Table \ref{LF_OII_TAB}. This is the first time the [O{\sc ii}] luminosity function has been measured at $z=2.2$, and also constitutes by far the largest sample of [O{\sc ii}] emitters at $z\sim2$. Since we do not constrain the faint-end slope, we fix it to the value measured at $z\sim1.5$ by \cite{Ly2007,Sobral12}, $\alpha=-0.9$.

We compare our results with results from the literature at lower redshifts: \cite{GalegoOII}, \cite{Takahashi07}, \cite{Ly2007}, \cite{Sobral12} and \cite{Bayliss}. Similarly to what has been found with H$\alpha$ and [O{\sc iii}]+H$\beta$, we find a significant L$^*$ evolution in the [O{\sc ii}] luminosity function. However, for [O{\sc ii}] we find an even stronger/faster L$^*$ evolution as L$^*$ becomes $\sim100\times$ brighter from the local Universe to $z=2.2$. We also show data from a recent study by \cite{Khostovan15}, based on the HiZELS (COSMOS+UDS) data. We find excellent agreement when comparing to \cite{Khostovan15}, which probes to slightly lower luminosities, while we probe a much larger volume. Indeed, our fit to SA22 fits the results from \cite{Khostovan15} well. However, even when combined, our data-sets are still unable to constrain the faint-end slope, and thus it is possible that the faint-end slope is steeper than $\alpha=-0.9$.

\subsection{Predictions for Euclid and WFIRST} \label{Euclid}

We find that approximately $\sim$50 per cent of the emitters in our sample (for our full sample down to our flux limit) are likely H$\alpha$ emitters at $z=0.81$, while [O{\sc iii}]+H$\beta$ account for $\sim18$ per cent and [O{\sc ii}] for about 9 per cent. The remaining $\sim23$ per cent consist of rarer emission lines (mostly at $z<0.8$, but also at $z>3$; see \citet{Matthee14}) and extra stars (with strong spectral features). If our flux limit was similar to the planned {\it Euclid} wide survey \citep{Laureijs12} of $>3\times10^{-16}$\,erg\,s$^{-1}$, we would obtain a lower fraction (but still significant) of line emitters which are not H$\alpha$, [O{\sc iii}]+H$\beta$ or [O{\sc ii}]: 18 \%. Our {\it Euclid}-like sample is dominated by H$\alpha$ emitters (67\%), followed by [O{\sc iii}]+H$\beta$ emitters (12\%), but still with some [O{\sc ii}] emitters (3\%). The raw (observed) density of H$\alpha$ emitters with $>3\times10^{-16}$\,erg\,s$^{-1}$ at $z\sim0.8$ ($\lambda\sim1.2$\,$\mu$m) is 8.66$\times10^{-5}$ Mpc$^{-3}$, while at the same wavelength the observed number density of [O{\sc iii}]+H$\beta$ emitters is about 5 times lower (1.6$\times10^{-5}$ Mpc$^{-3}$), and the number density of [O{\sc ii}] emitters (0.3$\times10^{-5}$ Mpc$^{-3}$) is almost 30 times lower than H$\alpha$ at $\lambda\sim1.2$\,$\mu$m. 

Using spectroscopic redshifts from the literature and from our own follow-up, we can have an even more robust quantification of the range of line emitters in the full sample. Figure \ref{PHOTOZs_spec} shows the full distribution of spectroscopic redshifts. Figure \ref{PHOTOZs_spec} also shows the different spectroscopic redshift sub-samples obtained when restricting our sample to a much higher narrow-band colour significance ($\Sigma$). For $\Sigma>3$ (our full sample), and based solely on spectroscopic redshifts, we find 59\% H$\alpha$ emitters, 9\% [O{\sc iii}]+H$\beta$ emitters, 7\% [S{\sc ii}] emitters, 4\% [O{\sc ii}] emitters, while 16\% are other $z<1.0$ (including [NII]) and the remaining 5\% are other $z>1$ emitters. Sub-samples with higher $\Sigma$ will be samples with higher flux limits, thus changing the distribution of emitters within such sub-samples. For our sample, $\Sigma>4$ corresponds to a cut in flux of $1.3\times10^{-16}$\,erg\,s$^{-1}$\,cm$^{-2}$, while the cut at $\Sigma>5$ corresponds to cutting the flux down to a flux limit of $1.55\times10^{-16}$\,erg\,s$^{-1}$\,cm$^{-2}$. For $\Sigma>4$ ($\Sigma>5$), based on spectroscopic redshift, we find 74\% (81\%) H$\alpha$ emitters, 9\% (7\%) [O{\sc iii}]+H$\beta$ emitters, 6\% (4\%) [S{\sc ii}] emitters and 3\% (3\%) [O{\sc ii}] emitters, while 6\% (3\%) are other $z<1.0$ (including [N{\sc ii}]) and the remaining 5\% (2\%) are other $z>1$ emitters. It is therefore clear that a selection with a higher flux limit will be more and more dominated by H$\alpha$ emitters, and thus our spectroscopic sample shows that both {\it Euclid} and {\it WFIRST} can expect $\sim80$\% of all line emitters to be H$\alpha$.


\subsection{Luminosity density and Cosmic star formation rate density}

By using the largest samples ever assembled and empirically computed uncertainties due to cosmic variance, we integrate our observed luminosity functions. We provide two different measurements for each emission line: i) the full analytical integral and ii) the numerical integration down to the observational limit. We provide the observed luminosity density values in Table \ref{SFRD_LFS}.

Prior to converting our observed luminosity densities for each line to a star formation rate density ($\rho_{\rm SFR}$), we apply two corrections. First we do a simple correction for dust extinction. We assume 1 magnitude at H$\alpha$ (A$_{\rm H\alpha}=1$). Dust corrections for [O{\sc iii}]+H$\beta$ and [O{\sc ii}] are discussed below. Secondly, we assume a 10-20\% AGN contribution to our samples, following our results in \S \ref{AGN}.

In order to convert luminosity densities to star formation rate densities we use, for H$\alpha$ \citep{Kennicutt}:

\begin{equation}
{\rm SFR}({\rm M}_{\odot} {\rm yr^{-1}})= 7.9\times 10^{-42} ~{\rm L}_{\rm H\alpha} ~ [{\rm erg\,s}^{-1}].
\end{equation}
For [O{\sc iii}]+H$\beta$, as seen before, directly interpreting as a star-formation indicator has a significant number of caveats. However, cautioning the reader that any values need to be interpreted with caution and are mostly indicative, we use the relation derived from \cite{OsterbrockFerland06}, in good agreement with the empirical calibration by \cite{Ly2007}, who found typical line ratios of $\sim1$:

\begin{equation}
{\rm SFR}({\rm M}_{\odot} {\rm yr^{-1}})= 7.35\times 10^{-42} ~{\rm L}_{\rm [OIII]+H\beta}~ [{\rm erg\,s}^{-1}].
\end{equation}
For [O{\sc ii}] we use the standard \cite{Kennicutt} calibration of [O{\sc ii}] as a star formation tracer (calibrated using H$\alpha$):

\begin{equation}
{\rm SFR}({\rm M}_{\odot} {\rm yr^{-1}})= 1.4\times 10^{-41} ~{\rm L}_{\rm [OII]}~ [{\rm erg\,s}^{-1}].
\end{equation}

For [O{\sc ii}], in order to correct for dust extinction, we use the results from \cite{Hayashi13} that present a large study of [O{\sc ii}] selected emitters at $z\sim1.47$ and use H$\alpha$ measurements for all of them (including stacking in H$\alpha$ for the faintest [O{\sc ii}] emitters) to calibrate [O{\sc ii}]. \cite{Hayashi13} shows that [O{\sc ii}] selected samples at much lower luminosities than ours have typical dust extinctions of A$_{\rm H\alpha}\approx0.4$, that their median extinction drops with increasing [O{\sc ii}] luminosity, and that assuming A$_{\rm H\alpha}=1.0$ for such emitters results in a significant overestimation of the real [O{\sc ii}] luminosity. As a simple test of this conclusion (and its validity at $z=2.2$), we start by assuming A$_{\rm H\alpha}=1.0$ for our sample of [O{\sc ii}] emitters at $z=2.2$, and find that we obtain $\rho_{\rm SFR}=1.0\pm0.3$\,M$_{\odot}$\,yr$^{-1}$ Mpc$^{-3}$, about 4-5 times higher than that given by H$\alpha$ \citep[][]{Sobral13}. This suggests that the results presented by \cite{Hayashi13} are also valid for $z=2.2$. Indeed, if we use their results and instead extinction-correct the [O{\sc ii}] luminosities assuming A$_{\rm H\alpha}\approx0.2$ (appropriate for our luminosity; this corresponds to A$_{\rm [O{\sc ii}]}\approx0.4$), we obtain a good agreement with the value from H$\alpha$ at the same redshift. However, we note that the uncertain faint-end slope of the [O{\sc ii}] luminosity function at high redshift also plays a role. At lower redshift, it is found to be relatively shallow ($\alpha$ approximately in the range from $-0.8$ to $-1.2$), and thus here we fix it to $\alpha=-0.9$, as this is the value found by the best studies at $z\sim1.5$ \citep[][]{Ly2007,Sobral12}.

While e.g. \cite{Ly2007} explored a double-blind [O{\sc iii}]-H$\alpha$ survey (and found typical line ratios around $\sim1$), no systematic investigation was performed in order to investigate the typical dust properties of [O{\sc iii}]+H$\beta$ emitters at $z>1$. However, as we have discussed, [O{\sc ii}]-selected emitters are significantly less dusty than H$\alpha$ selected line emitters (as an [O{\sc ii}]-selection preferentially picks up dust-free sources and is biased against dusty sources). This is likely to be the case for [O{\sc iii}]+H$\beta$ emitters, although we expect them to present typical dust extinction properties which are more similar to H$\alpha$-selected line emitters. However, given that we are using a SFR calibration based directly on observed line fluxes which used the H$\alpha$ line \cite[][]{Ly2007}, we will use the correction appropriate for H$\alpha$-selected emitters, A$_{\rm H\alpha}=1$, corresponding to about 1.5 mag at [O{\sc iii}]+H$\beta$. We obtain $\rho_{\rm SFR}=0.14\pm0.03$\,M$_{\odot}$\,yr$^{-1}$ Mpc$^{-3}$.

\begin{figure}
\centering
\includegraphics[width=8cm]{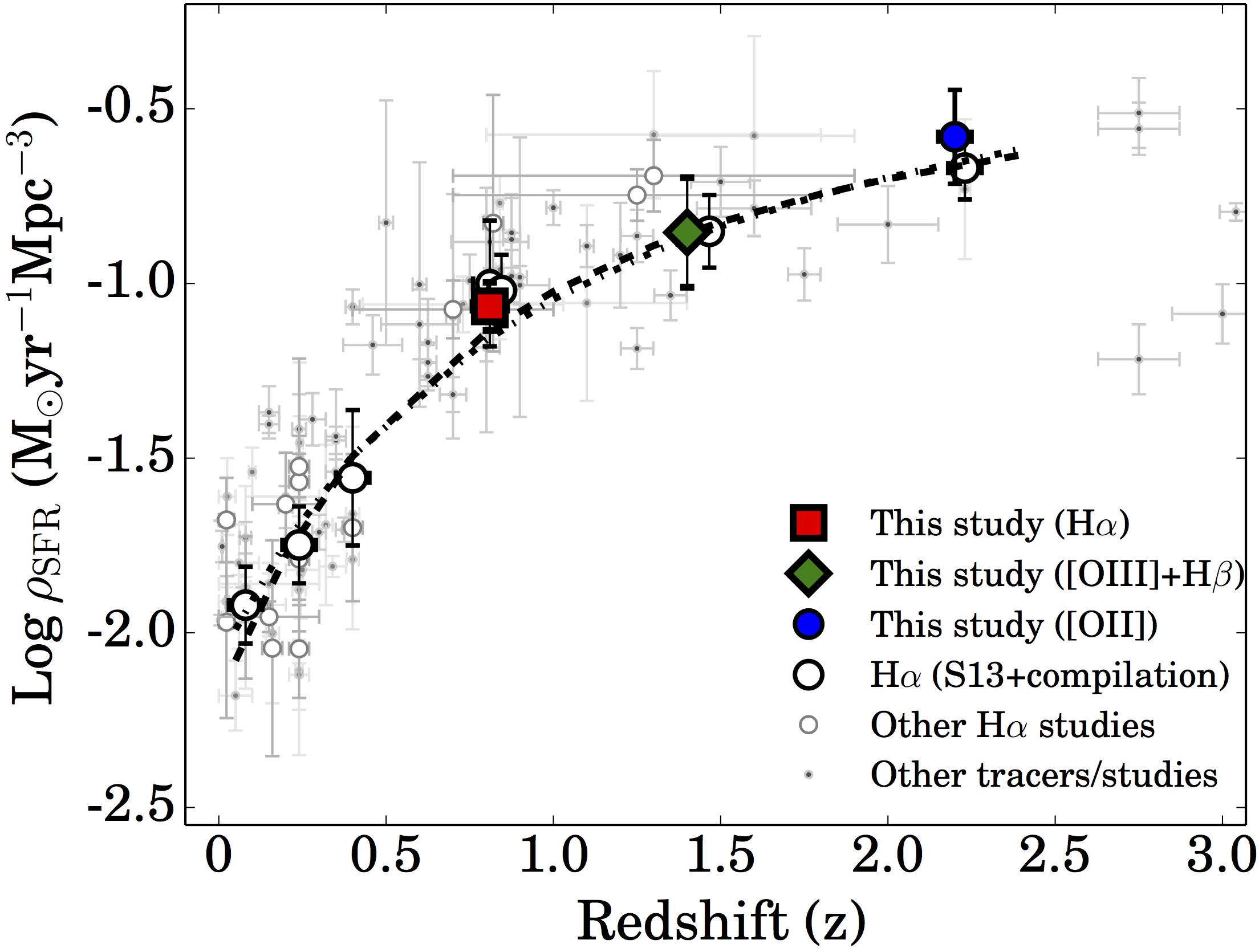}
\caption{Star formation history of the Universe using our CF-HiZELS SA22 data: $z=0.81$ (H$\alpha$), $z=1.4$ ([O{\sc iii}]+H$\beta$) and $z=2.2$ ([O{\sc ii}]). Our results agree very well with the H$\alpha$ star-formation history of the Universe \citep[dashed line is the parameterisation from][]{Sobral13}.} 
\label{SFRH}
\end{figure}

Our results from H$\alpha$, [O{\sc iii}] and [O{\sc ii}] can be found in both Table \ref{SFRD_LFS} and in Figure \ref{SFRH}. We find a very good agreement with the overall star formation history of the Universe \citep[e.g.][]{Karim,Sobral13}, indicating that our simple assumptions for dust and AGN correction (or the combination of both) provide decent average estimates for the full population of galaxies \citep[although it obviously fails on a source by source basis, and it will have strong dependences with mass, luminosity/etc -- see e.g. ][]{Ibar13,Sobral14}. 

Our results confirm the rise of the star formation history up to $z=2.2$, but also indicate how important it is to apply appropriate dust extinction corrections \citep{Hayashi13} that take into account that [O{\sc ii}] emitters are typically almost dust free. While there are significant uncertainties in the use of [O{\sc ii}] as a star formation indicator for higher redshifts where H$\alpha$ is not available, if a better understanding of the typical dust properties is obtained, and particularly if the faint-end slope is constrained, it should provide at least a competitive view when compared to e.g. the UV.

%
%
%
%

\subsection{Over-densities}

\begin{figure}
\centering
\includegraphics[width=8cm]{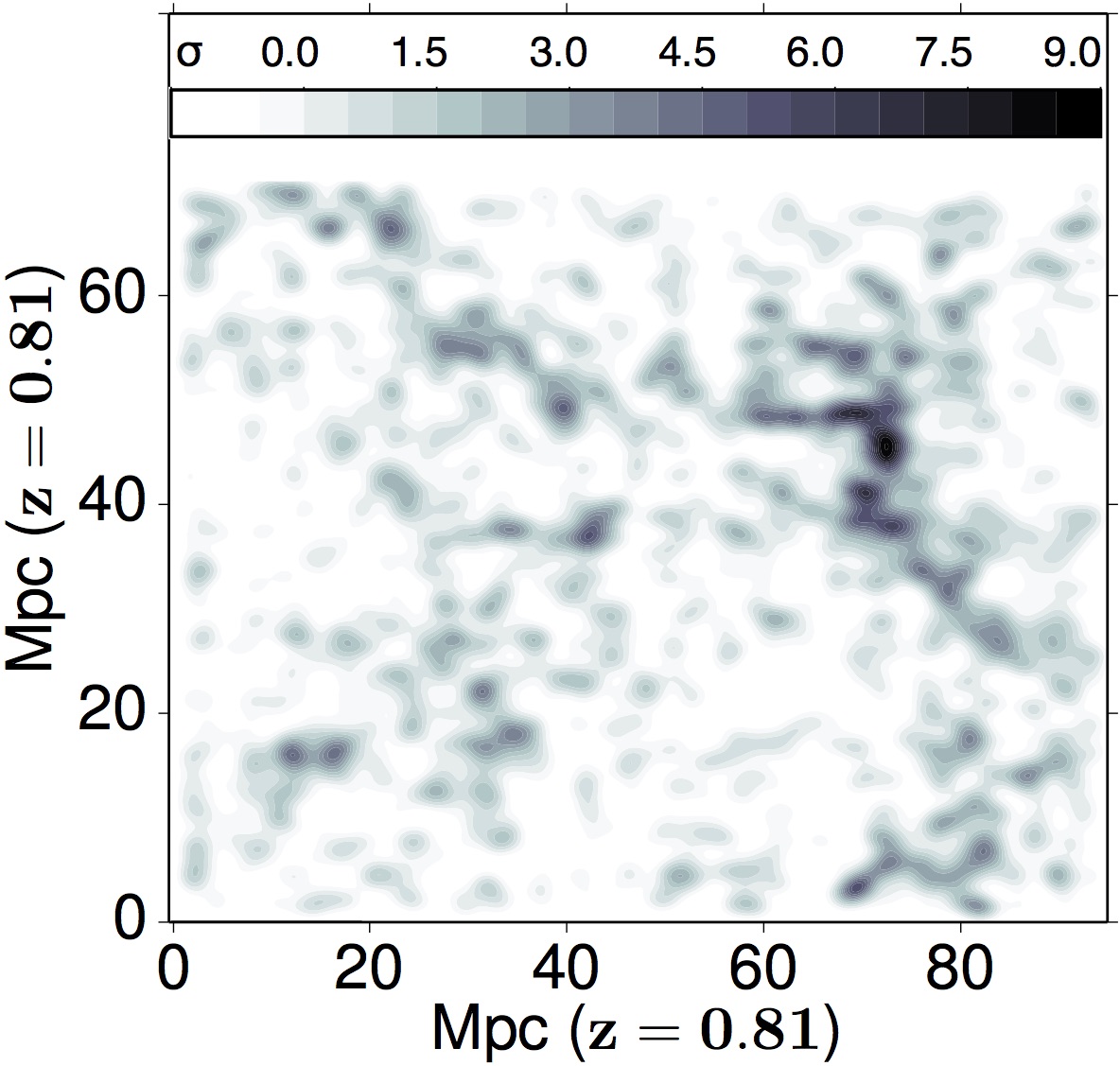}
\caption{\small{Density of H$\alpha$ emitters at $z = 0.81$ in our SA22 survey. The angular scale is converted to a physical scale. The smoothed distribution highlights the large 7-9$\sigma$ over-density, which is approximately 15 Mpc across. It also shows the filamentary structure of the large scale structure in the Universe.}} 
\label{fig:radechalpha}
\end{figure}

We explore our very wide area coverage over different fields to look for significant over-densities. In order to do this, we take two approaches: we smooth the distribution of sources, but also compute the 10-th nearest neighbour densities. As can be seen in Figure \ref{fig:radechalpha}, there appears to be a significant large-scale over-density of H$\alpha$ emitters which contains $\sim $\,300 candidate $z=0.81$ H$\alpha$ line emitters within a $\sim $\,20-arcmin field (Figure ~\ref{fig:radechalpha}). This includes a region where the number density of H$\alpha$ emitters is $\sim10$ times higher than the general field. This strong over-density of H$\alpha$ emitters in the SA22 field has now been confirmed with the new KMOS instrument \citep[][]{Sobral13b}. This $\sim8\sigma$ over-density of H$\alpha$ emitters is found within a volume of  3000\,Mpc$^3$ (co-moving).

%
%
%
%

We note that this finding could be important to also interpret any over-density of star-forming galaxies detected at this or at even higher redshift. Without a careful analysis of the full galaxy population, one could be misled to conclude that such structure could be a cluster and/or a proto-cluster. However, following \cite{Sobral11}, \cite{Sobral13b} find that the structure presents galaxy number densities which are more typical of groups, or intermediate densities, but not clusters. In practice, the structure is likely a dense cosmic web structure or filamentary/wall-like as a group. \cite{Darvish14} find that the fraction of star-forming galaxies (H$\alpha$ emitters) is much higher in filaments than in the general field or in clusters, making these easily detectable with wide H$\alpha$ surveys.

We also search for over-densities within the samples of [O{\sc iii}] and [O{\sc ii}] emitters, but find only mild over-densities ($<3$\,$\sigma$). Nevertheless, these density fluctuations still result in significant cosmic variance, if only different parts of the fields would be investigated.

%
%
%
\begin{figure}
\centering
\begin{tabular}{c}
\includegraphics[width=8.25cm]{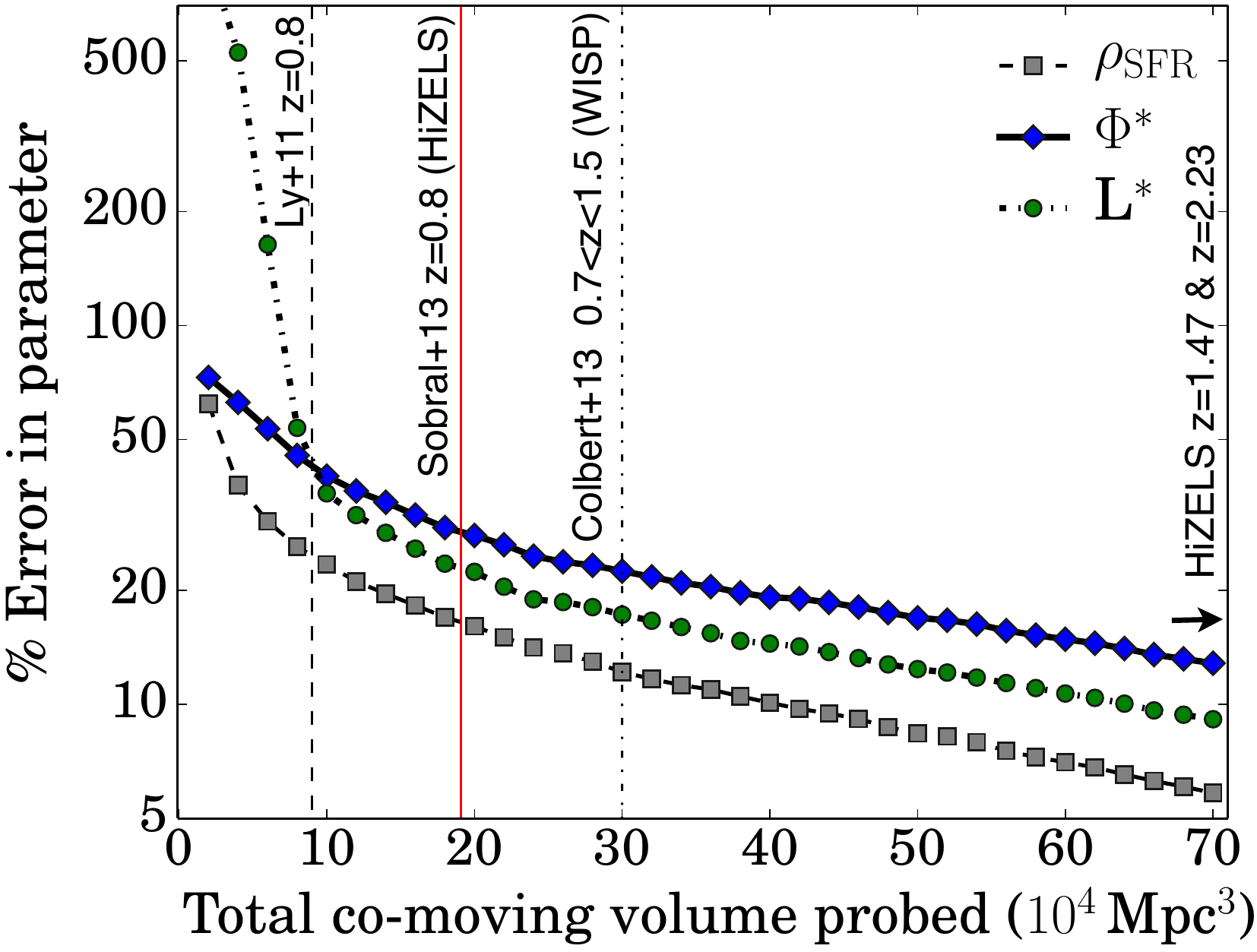} \\
\includegraphics[width=8.25cm]{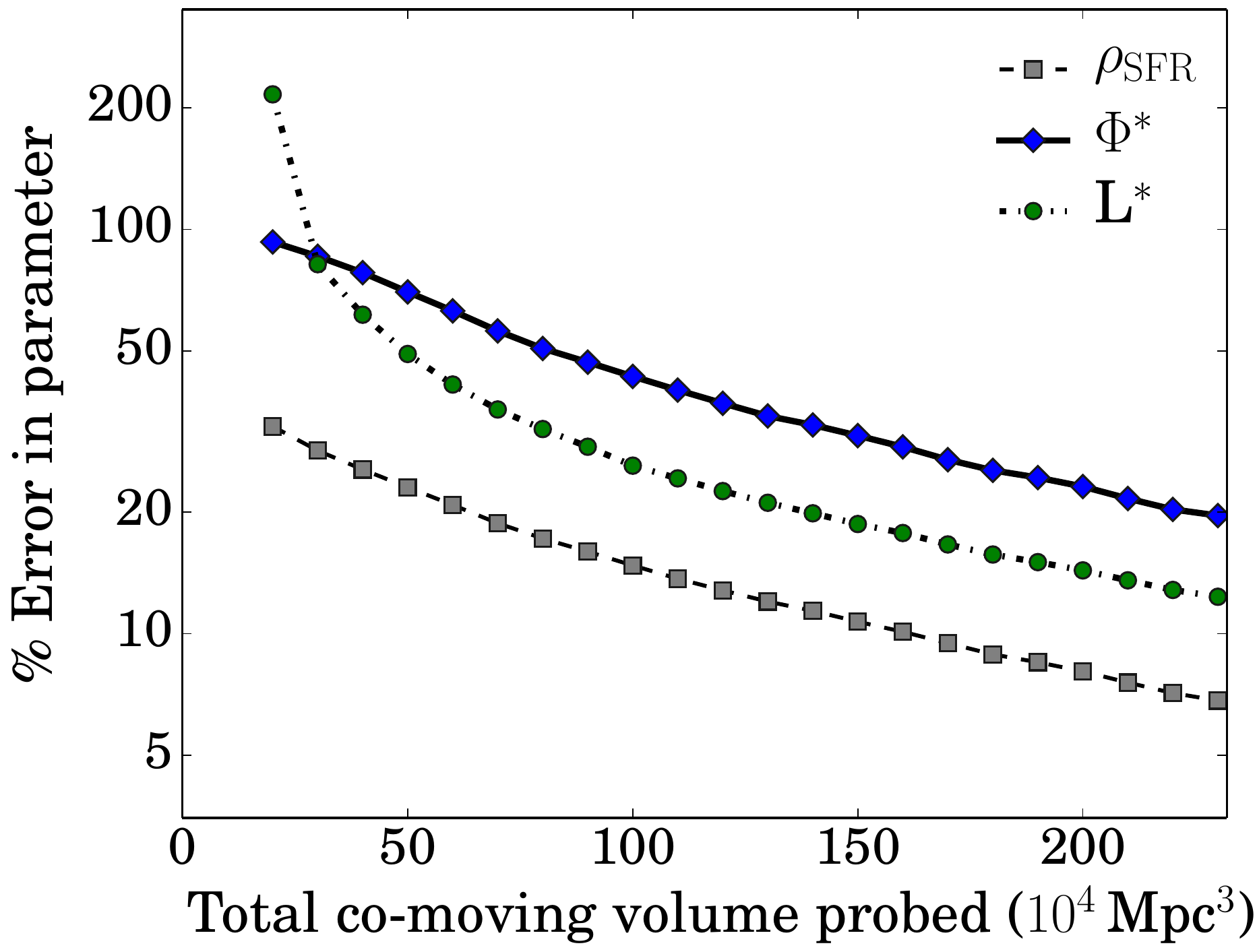} \\
\includegraphics[width=8.25cm]{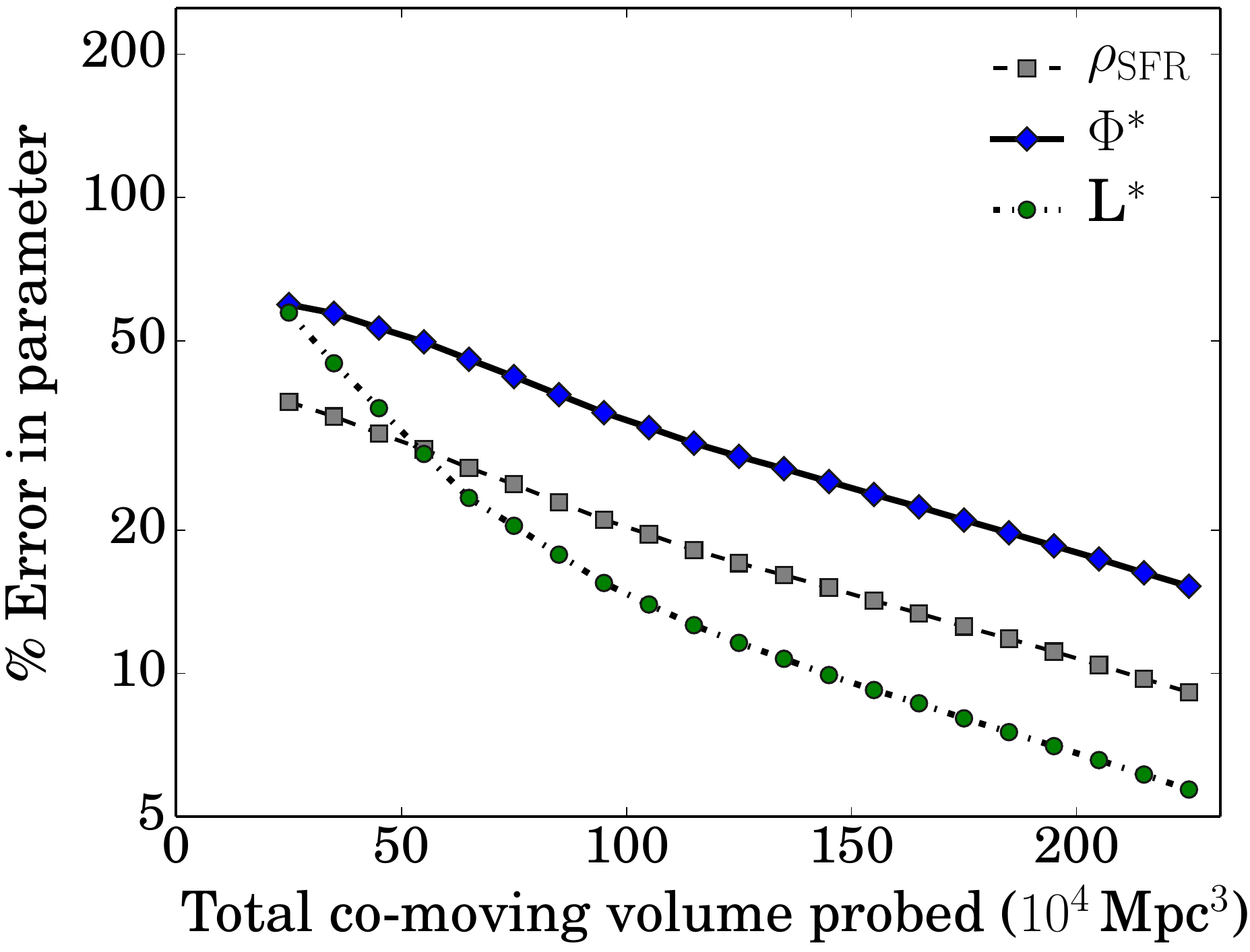}\\
\end{tabular}
\caption{{\it Top:} The error/variance on $ \phi^*$, L$^*$ and SFRD ($\rho_{\rm SFR}$) as function of total probed co-moving volume, based on a million realisations, each sampling a random combination of UDS, COSMOS and SA22 for H$\alpha$ LF at $z\sim0.8$. The results show that the variance on all parameters/quantities decreases with total co-moving volume probed, but that it is significant for most volumes usually used in the literature. Our results are therefore expected to have an error $<10$\%. {\it Middle:} The error/variance on $ \phi^*$, L$^*$ and SFRD ($\rho_{\rm SFR}$) as function of total probed co-moving volume, based on a million realisations, each sampling a random combination of UDS, COSMOS and SA22 for [O{\sc iii}]+H$\beta$ LF at $z\sim1.4$. {\it Bottom:} The error/variance on $ \phi^*$, L$^*$ and SFRD ($\rho_{\rm SFR}$) as function of total probed co-moving volume, based on a million realisations, each sampling a random combination of UDS, COSMOS and SA22 for [O{\sc ii}] LF at $z\sim2.2$.} 
\label{cosmic_Ha}
\label{cosmic_OIII}
\label{cosmic_OII}
\end{figure}

\subsection{Cosmic Variance and Cosmic convergence}

We explore the large samples spread over large, multiple areas (SA22, COSMOS, UDS) to empirically quantify cosmic variance affecting narrow-band surveys. In order to do this, we follow a similar procedure to \cite{Sobral11}, and divide the full sample in several areas corresponding to individual pointing/cameras, roughly corresponding to 0.02 to 0.05 deg$^2$. We then repeat the determination of the luminosity functions by randomly selecting these areas. We start by computing them for the smallest, contiguous 0.02 deg$^2$ areas and then go up by steps of 0.02, sampling these in random combinations within SA22, COSMOS and UDS. We first investigate how $\alpha$, $ \phi^*$ and $L^*$ are affected by sample variance, but find that, due to the depth of our data, $\alpha$ is in general completely unconstrained. Thus, for the following analysis, we fix $\alpha$ and concentrate on studying $ \phi^*$ and $L^*$ only. For each emission line, we obtain a total of a million realisations of the luminosity function, spread over the minimum to the maximum total area. Quoted errors (in \%) are computed as the ratio between the standard deviation of a parameter for each bin in area (from all LF realisations with a given total area) and the actual parameter value.

We note that our results can have a relatively general application for data-sets with similar depths, but that for data-sets that are e.g. significantly shallower, the errors measured with our method would become much larger, as they would be dominated by e.g. Poissonion errors and, in such cases (e.g. if depth is shallower than $L^*$), $ \phi^*$ and L$^*$ become unconstrained, and thus the results of our study no longer apply (in order to lower the errors in the parameters, deeper data is more efficient than probing larger areas). We also note that for [O{\sc iii}]+H$\beta$ and [O{\sc ii}] LFs, due to the depth of our data, we only probe down to $\sim L^*$ and thus, for the lowest volumes, our errors become completely Poissonion dominated, resulting in very large variance which is due to low number statistics, not cosmic variance. Thus, we do not show the results for areas which are so small that low number statistics completely dominate.

The results are presented in Figure \ref{cosmic_OII}. As expected, we find that cosmic/sample variance is very strong for small volumes ($<10^5$\,Mpc$^3$), and strongly reduced with increasing area/volume. For the H$\alpha$ luminosity function, for example, a total area of $\approx1$\,deg$^2$ ($\approx10^5$\,Mpc$^3$) still results in significant scatter/significant errors due to sample/cosmic variance, where $L^*$ is clearly the most affected parameter (up to $>$100\% errors), followed by $ \phi^*$ ($\sim40-50$\% errors).

With our total area of $\sim9$\,deg$^2$ the errors on $ \phi^*$ are reduced to $\sim10-15$\%, while the errors on $L^*$ drop to $\sim8$\% and the error on SFRD drops to close to 5\%. We therefore conclude that only by probing volumes comparable or larger than 10$^6$\,Mpc$^3$ will the sample variance fully drop below 10\%. Our results are also very useful to interpret results from similar previous, current and future surveys of star-forming galaxies, particularly to interpret differences in number counts and luminosity functions, without having to rely on simulations and on other indirect methods/arguments.

We find similar results for both [O{\sc ii}] and [O{\sc iii}]+H$\beta$. We note that we use our results to obtain a better estimate of our errors by adding (in quadrature) the expected errors due to cosmic variance, based on the total volume probed by each luminosity function/measurement. This is a small contribution to the total error budget given that our survey probes a very large volume, but would be a significant amount to surveys probing small volumes.

%
%
\begin{table*}
 \centering
  \caption{The luminosity function and star formation rate density evolution for $z=0.8,1.4,2.2$. L$^*$, $ \phi^*$ and $\alpha$ values are derived without any correction for dust extinction. The measurements corrected for 1 mag extinction at H$\alpha$ for H$\alpha$ and [O{\sc iii}]+H$\beta$ emitters (as our SFR calibration is based on observed H$\alpha$ and [O{\sc iii}] line fluxes, thus the correction for H$\alpha$ emitters is the most appropriate), but follow \citet{Hayashi13} for [O{\sc ii}] selected emitters (A$_{\rm H\alpha}=0.2$). Columns present the redshift, break of the luminosity function, $L^*_{\rm line}$, normalisation ($ \phi^*_{\rm line}$) and faint-end slope ($\alpha$) of the luminosity functions. The two right columns present the star formation rate density at each redshift based on integrating the luminosity function down to each observational limit (41.4 for H$\alpha$, 42.1 for [O{\sc iii}] and 42.6 for [O{\sc ii}] in logL) and for a full integration, and include dust corrections. The two columns immediately to the left present similar measurements but for luminosity densities. Star formation rate densities include a correction for AGN contamination of 10\% at $z=0.8$ for H$\alpha$ \citep[see][]{Garn2010a}, up to $L^*$, and a correction given by $0.59\times\log_{10}(L/L^*)+0.112$ for $L>L^*$ \citep[][]{Sobral15}. For [O{\sc iii}]+H$\beta$ at $z\sim1.4$, we assume an AGN contamination of 15\%, while we assume a contamination of 20\% at $z=2.2$ for [O{\sc ii}] emitters. We only fit $\alpha$ for the H$\alpha$ LF and fix it for all the other luminosity functions. $L^*$ and $ \phi^*$ are obtained by fixing $\alpha=-1.6$ (H$\alpha$), and the 1\,$\sigma$ errors on $L^*$ and $ \phi^*$ are derived from such fits (with fixed $\alpha$). }
  \begin{tabular}{@{}ccccccccc@{}}
  \hline
   ($z$) Em. Line & $L^*$ & $ \phi^*$ & $\alpha$ & log\,$\rho_{L}$ obs  & log\,$\rho_{L}$  & $ \rho_{\rm SFR}$  obs  & $ \rho_{\rm SFR}$ All  \\
     ($z\pm0.02$)    & (erg\,s$^{-1}$) & (Mpc$^{-3}$) &    & (erg\,s$^{-1}$\,Mpc$^{-3}$) & (erg\,s$^{-1}$\,Mpc$^{-3}$) & (M$_{\odot}$\,yr$^{-1}$ Mpc$^{-3}$)   & (M$_{\odot}$\,yr$^{-1}$ Mpc$^{-3}$)  \\
 \hline
   \noalign{\smallskip}
($z=0.81$) \ H$\alpha$ & $41.72^{+0.03}_{-0.02}$ & $-2.31^{+0.04}_{-0.05}$ & $-1.6^{+0.2}_{-0.2}$ & $38.17^{+0.03}_{-0.03}$ & $40.14^{+0.01}_{-0.01}$ & $0.022^{+0.002}_{-0.002}$  & $0.086^{+0.003}_{-0.003}$ \\
($z=1.37$) $\bf \rm {[OIII]}$ & $42.10^{+0.05}_{-0.04}$ & $-2.71^{+0.08}_{-0.09}$ & $-1.6$ (Fixed) & $40.15^{+0.06}_{-0.02}$ & $40.36^{+0.05}_{-0.04}$ & $0.09^{+0.01}_{-0.01}$  & $0.14^{+0.04}_{-0.04}$ \\  
($z=2.20$) $\bf \rm {[OII]}$ & $42.23^{+0.05}_{-0.04}$ & $-2.23^{+0.14}_{-0.14}$ & $-0.9$ (Fixed) & $39.01^{+0.01}_{-0.01}$ & $40.02^{+0.10}_{-0.10}$ & $0.11^{+0.02}_{-0.02}$  & $0.26^{+0.03}_{-0.04}$ \\
 \hline
\end{tabular}
\label{SFRD_LFS}
\end{table*}

%


\section{Conclusions}

We presented results from the largest contiguous narrow-band survey in the near-infrared ($J$ band). We surveyed $\approx$10\,deg$^2$ of contiguous extragalactic sky and found a total of 5976 robust candidate emission-line galaxies ($\Sigma>3$). By using deep CFHTLS $ugriz$ and UKIDSS DXS $J$ and $K$, a large sample of available spectroscopic redshifts from VVDS and VIPERS from the literature, and by obtaining new spectroscopic follow-up observations with MOSFIRE and FMOS, and by combining our sample with HiZELS, we derive the largest samples of emission-line selected galaxies. We find by far the largest sample of H$\alpha$ emitters at $z\sim0.8$: 3471 sources ($\sim400$ spectroscopically confirmed), obtaining the most accurate measurement of the star formation rate density at that cosmic epoch. We also present the largest sample (1341) of [O{\sc iii}]/H$\beta$ emitters at $z\sim1.4$ and present the first large sample (572 sources) of  [O{\sc ii}] emitters at the peak of the star formation history ($z=2.2$). Our main conclusions are:

\begin{itemize}

\item Our large spectroscopic sample from FMOS and MOSFIRE allows us to confirm that the [N{\sc ii}]/H$\alpha$ correction as a function of EW(H$\alpha$+[N{\sc ii}]) from SDSS is applicable to $z=0.81$ with no evolution. We nonetheless provide a simpler linear fit, based on our data, that can be used to correct similar data-sets: $f(\rm [NII]/H\alpha)=-0.296\times\log_{10}(\rm EW_{\rm H\alpha+[NII]})+0.8$. Our spectroscopic H$\alpha$ sample also shows that our H$\alpha$ emitters have a metallicity of $\rm 12+log_{10}(O/H)=8.56\pm0.05$, slightly sub-solar. We also make significant detections of the [S{\sc ii}] doublet, with a line ratio of I([S{\sc ii}]${6716}$)/I([S{\sc ii}]${6731}) = 1.33 \pm 0.08$, implying an electron density of  40-200\,cm$^{-3}$. The [S{\sc ii}]${6716}$/I(H$\alpha$)$=0.14\pm0.02$ ratio also implies ionisation potential of log$_{10}{\rm U}=-3.9\pm0.5$\,cm$^{-2}$.

\item We obtain H$\alpha$, [O{\sc iii}]+H$\beta$ and [O{\sc ii}] luminosity functions at $z=0.8,1.4,2.2$ with the largest statistical samples, reaching up to the highest luminosities. We find a strong luminosity evolution ($L^*$) in the luminosity function of all the lines with increasing redshift up to at least $z\sim2.2$, with a less significant (but present) $ \phi^*$ evolution. This is consistent with the evolution seen across redshift presented by \cite{Khostovan15}.

\item We show that the [O{\sc iii}]+H$\beta$ luminosity function at $z=1.4$ is very hard to interpret in general, particularly due to the complicated contribution from the two different [O{\sc iii}] lines, and, to a lesser extent, to the contribution of H$\beta$. While we find that [O{\sc iii}]${5007}$ (only) dominates the sample of spectroscopically confirmed [O{\sc iii}]+H$\beta$ emitters, representing $\sim50$\% of the full sample, [O{\sc iii}]${4959}$ (only) represents 27\%, almost twice as common as H$\beta$. We find that 7\% of the [O{\sc iii}]+H$\beta$ sources are actually detections of both [O{\sc iii}] lines at opposite wings of the filter. 

\item We present the first [O{\sc ii}] luminosity function at $z=2.2$, and find a very strong evolution from $z\sim0$ to $z\sim2.2$, much stronger than that seen for either H$\alpha$ or [O{\sc iii}]+H$\beta$. By correcting for dust extinction using \cite{Hayashi13} the star formation rate density based on [O{\sc ii}] is in excellent agreement with H$\alpha$ \citep{Sobral13}. If 1 magnitude of extinction at H$\alpha$ was used for [O{\sc ii}] emitters instead, the star formation rate density would have been over-estimated by a factor of 2-3. Thus, if the \cite{Hayashi13} calibration is used (and if it remains valid for even higher redshift), [O{\sc ii}] may be a reasonably good way to measure star formation rate density beyond $z\sim2.2$, as H$\alpha$ gets redshifted out of the $K$ band.

\item We find a reasonable good agreement between the star formation rate density from [O{\sc iii}] and that from H$\alpha$ at a similar redshift. We nonetheless caution that without a detailed investigation into the nature of [O{\sc iii}]+H$\beta$ emitters (and without robustly separating them from H$\beta$ emitters, interesting on their own), using [O{\sc iii}]+H$\beta$ as a star formation indicator at high redshift is highly unreliable.

\item For the planned {\it Euclid} wide survey \citep{Laureijs12} flux limit of $>3\times10^{-16}$\,erg\,s$^{-1}$, our sample is dominated by H$\alpha$ emitters (67\%), followed by [O{\sc iii}]+H$\beta$ emitters (12\%), but still with some [O{\sc ii}] emitters (3\%); the remaining 18\% are rarer emitters spread over a wide range of redshifts. The raw (observed) density of H$\alpha$ emitters with $>3\times10^{-16}$\,erg\,s$^{-1}$ at $z\sim0.8$ ($\lambda\sim1.2$\,$\mu$m) is 8.66$\times10^{-5}$ Mpc$^{-3}$, while at the same wavelength the observed number density of [O{\sc iii}]+H$\beta$ emitters is about 5 times lower (1.6$\times10^{-5}$ Mpc$^{-3}$), and the number density of [O{\sc ii}] emitters (0.3$\times10^{-5}$ Mpc$^{-3}$) is almost 30 times lower than H$\alpha$ at $\lambda\sim1.2$\,$\mu$m. Our fully spectroscopically confirmed sample confirms these numbers, predicting that $\sim70-80$\% of all line emitters found at $\lambda\sim1.2$\,$\mu$m will be H$\alpha$.

\item We find significant over-densities, out of which the strongest one is found at $z=0.8$ and traced by H$\alpha$ emitters. It is an $8.5$\,$\sigma$ (confirmed with KMOS) over-density of H$\alpha$ emitters, where the number density is a factor $\sim10\times$ higher than the average; this is consistent with group-like densities, and most likely a rich filamentary structure, similar to what has been studied in \cite{Darvish14}. Only mild over-densities are found for [O{\sc iii}]+H$\beta$ and [O{\sc ii}] emitters, although such over-densities would be much harder to find given the relatively shallow data when compared to H$\alpha$.

\item We take advantage of the large volumes/area and multiple fields to sub-divide the samples in randomised areas and provide a robust empirical measurement of sample/cosmic variance for the different lines/redshifts. We find that surveys for star-forming/emission-line galaxies can only overcome cosmic-variance (errors $<10$\%) if they are based on volumes $>5\times10^{5}$\,Mpc$^{3}$. In other words, multiple/different volumes adding up to $<5\times10^{5}$\,Mpc$^{3}$ show variance which results in errors being $>10$\%.

\item  Errors due to sample (cosmic) variance on surveys probing $\sim10^4$\,Mpc$^{3}$ and $\sim10^5$\,Mpc$^{3}$ are typically very high: $\sim300$\% and $\sim40-60$\%, respectively. Focusing only on $L^*$ and $ \phi^*$, the latter is the most affected parameter for large volumes, while $L^*$ is completely un-determined for volumes $<10^5$\,Mpc$^{3}$. 

\end{itemize}

\section*{Acknowledgments}

The authors wish to thank the anonymous reviewer for many helpful comments and suggestions which greatly improved the clarity and quality of this work. DS acknowledges financial support from the Netherlands Organisation for Scientific research (NWO) through a Veni fellowship, from FCT through a FCT Investigator Starting Grant and Start-up Grant (IF/01154/2012/CP0189/CT0010), from FCT grant PEst-OE/FIS/UI2751/2014, and from LSF and LKBF. JM acknowledges the award of a Huygens PhD fellowship. PNB is grateful for support from STFC. IRS acknowledges support from STFC, a Leverhulme Fellowship, the ERC Advanced Investigator programme DUSTYGAL and a Royal Society/Wolfson Merit Award. BMJ acknowledges support from the ERC-StG grant EGGS-278202. The Dark Cosmology Centre is funded by the DNRF. The Dark Cosmology Centre is funded by the DNRF. JWK acknowledges support from the National Research Foundation of Korea (NRF) grant, No. 2008-0060544, funded by the Korea government (MSIP). JPS acknowledges support from STFC (ST/I001573/1). JC acknowledges support from the FCT-IF grant IF/01154/2012/CP0189/CT0010. The work was only possible due to OPTICON/FP7 and the invaluable access that it granted to the CFHT telescope. We would also like to acknowledge the excellent work done by CFHT staff in conducting the observations in service mode, and on delivering truly excellent data. We are also tremendously thankful to Kentaro Aoki for the incredible support while observing at Subaru with FMOS, and also to the Keck staff for the help with the observations with MOSFIRE. This work is based on observations obtained with WIRCam on the CFHT, OPTICON programme 2011B/029, 2012A019 and 2012B/016. Based on observations made with ESO telescopes at the La Silla Paranal Observatory under programmes IDs 60.A-9460 (data can be accessed through the ESO data archive), 087.A-0337 and 089.A-0965. Based on observations done with FMOS on Subaru under program S14A-084, and on MOSFIRE/Keck observations under program U066M. Part of the data on which this analysis is based are available from \cite{Sobral13}. Dedicated to the memory of C. M. Sobral (1953-2014).

\bibliographystyle{mn2e.bst}
\bibliography{bibliography.bib}

\appendix

\section{Catalogue of CF-HiZELS narrow-band emitters}  \label{Cats}

The catalogue of narrow-band emitters in SA22 is presented in Table A.1. It contains IDs, Right Ascension (RA, J2000), Declination (Dec, J2000), narrow-band magnitude (NB, AB), broad-band magnitude ($J$, AB), the significance of the narrow-band excess ($\Sigma$, estimated in 2$''$ apertures), estimated flux ($\log_{10}$), estimated observed EW (\AA), and a flag for those that are classified as H$\alpha$ at $z=0.8$ (1), [O{\sc iii}] or H$\beta$ at $z\sim1.4$ (2) and [O{\sc ii}] at $z=2.20$ (3). Unclassified sources as flagged with 0 and candidate stars are identified with flag -1. We note that there are a few (15) sources with some extreme EWs (observed EWs $>10^4$\,\AA \ [observed]) and 17 sources with $\Sigma>100$, but that these may be supernovae and/or strongly variable sources \citep[see e.g.][]{Matthee2015}, apart from real extremely rare sources. Note that only the online version contains the full catalogue -- here only three entries of the table are shown as examples of the entire catalogue.

%
%
\begin{table*}
 \centering
  \caption{Example entries form the catalogue of all $\Sigma>3$ narrow-band sources selected in the SA22 field from this paper. The full catalogue is available on-line.}
  \begin{tabular}{@{}ccccccccccc@{}}
  \hline
   ID & R.A. & Dec. & NB$_J$ & $J$ & $\Sigma$  & log Flux  & EW$_{\rm obs}$ & Class. Flag  \\
         & {(J2000)} &(J2000) & (AB) &(AB) &  & erg\,s$^{-1}$\,cm$^{-2}$ & \AA\\
 \hline
   \noalign{\smallskip}
CFHIZELS-SM14-99905 & 22\, 09\, 11.64 & $+$01\,23\,13.25 & 20.43$\pm0.06$ & 21.40$\pm0.06$ & 10.1 & $-15.48$ & 170.8 & 1 & \\ 
CFHIZELS-SM14-98588 & 22\,19\, 34.55 & $+$00\, 24\, 51.36& 20.41$\pm0.06$ & 22.24$\pm0.11$ & 13.2 & $-15.33$ & 657.2 & 2 & \\
CFHIZELS-SM14-97547 & 22\, 11\, 31.53 & $-$00\, 37\, 16.69 & 20.40$\pm0.07$ & 20.82$\pm0.05$ & 5.1 & $-15.72$ & 52.4 & 3 & \\
 \hline
\end{tabular}
\label{NBJ_CAT}
\end{table*}

\section{Luminosity functions}

%
%
\begin{table}
 \centering
  \caption{H$\alpha$ Luminosity Function at $z=0.81$ from our SA22 survey. Note that L$_{\rm \bf H\alpha}$ has been corrected for [N{\sc ii}] contamination, but not for dust extinction. The total volume per bin is $8.6\times10^5$\,Mpc$^3$, assuming a top hat filter, but $ \phi$ corr has been corrected for both incompleteness and the fact that the filter profile is not a perfect top hat.}
  \begin{tabular}{@{}cccc@{}}
  \hline
  \bf $\bf \log$\bf L$_{\rm \bf H\alpha}$ & \bf Sources &$\bf  \phi$ \bf obs & $\bf  \phi$ \bf corr   \\
 \hline
 \bf  $\bf z=0.81$ & \# & Mpc$^{-3}$ & Mpc$^{-3}$ \\
  \hline
   \noalign{\smallskip}
$41.40\pm0.025$  & $445\pm21$ & $-1.98\pm0.02$ & $-2.01\pm0.02$   \\
$41.45\pm0.025$  & $433\pm20$ & $-1.99\pm0.02$ & $-2.03\pm0.02$  \\
$41.50\pm0.025$  & $370\pm19$ & $-2.06\pm0.02$ & $-2.09\pm0.02$  \\
$41.55\pm0.025$  & $322\pm17$ & $-2.12\pm0.02$ & $-2.15\pm0.03$  \\
$41.60\pm0.025$  & $268\pm16$ & $-2.20\pm0.03$ & $-2.23\pm0.03$  \\
$41.65\pm0.025$  & $248\pm15$ & $-2.24\pm0.03$ & $-2.27\pm0.03$  \\
$41.70\pm0.025$  & $201\pm14$ & $-2.33\pm0.03$ & $-2.35\pm0.03$  \\
$41.75\pm0.025$  & $159\pm12$ & $-2.43\pm0.04$ & $-2.46\pm0.04$   \\
$41.80\pm0.025$  & $109\pm10$ & $-2.59\pm0.04$ & $-2.63\pm0.05$   \\
$41.85\pm0.025$  & $76\pm8$ & $-2.75\pm0.05$ & $-2.79\pm0.06$   \\
$41.90\pm0.025$  & $73\pm8$ & $-2.77\pm0.05$ & $-2.81\pm0.06$  \\
$41.95\pm0.025$  & $47\pm6$ & $-2.95\pm0.07$ & $-2.98\pm0.07$  \\
$42.00\pm0.025$  & $46\pm6$ & $-2.96\pm0.07$ & $-3.01\pm0.07$  \\
$42.05\pm0.025$  & $23\pm4$ & $-3.26\pm0.10$ & $-3.31\pm0.10$  \\
$42.10\pm0.025$  & $21\pm4$ & $-3.29\pm0.10$ & $-3.33\pm0.11$  \\
$42.15\pm0.025$  & $13\pm3$ & $-3.50\pm0.14$ & $-3.54\pm0.14$  \\
$42.20\pm0.025$  & $9\pm3$ & $-3.64\pm0.17$ & $-3.68\pm0.17$  \\
$42.25\pm0.025$  & $5\pm2$ & $-3.93\pm0.26$ & $-4.00\pm0.26$   \\
$42.30\pm0.025$  & $5\pm2$ & $-3.93\pm0.26$ & $-3.93\pm0.26$  \\
$42.45\pm0.025$  & $13\pm3$ & $-4.44\pm0.14$ & $-4.44\pm0.14$  \\
$42.65\pm0.025$  & $3\pm1$ & $-5.08\pm0.33$ & $-5.02\pm0.33$  \\
\end{tabular}
\label{LF_HA_TAB}
\end{table}

%
%
\begin{table}
 \centering
  \caption{[OIII]+H$\beta$ Luminosity Function -- this has not been corrected for extinction. The total volume per bin is $24.8\times10^5$\,Mpc$^3$, assuming a top hat filter, but $ \phi$ corr has been corrected for both incompleteness and the fact that the filter profile is not a perfect top hat.}
  \begin{tabular}{@{}cccc@{}}
  \hline
  \bf $\bf \log$\bf L$_{\rm \bf [OIII]}$ & \bf Sources &$\bf  \phi$ \bf obs & $\bf  \phi$ \bf corr \\
 \hline
 \bf  $\bf z=1.4$ & \# & Mpc$^{-3}$ & Mpc$^{-3}$  \\
  \hline
   \noalign{\smallskip}
$42.07\pm0.03$  & $285\pm16$ & $-2.72\pm0.03$ & $-2.71\pm0.04$ \\
$42.13\pm0.03$  & $283\pm16$ & $-2.72\pm0.03$ & $-2.71\pm0.04$   \\
$42.19\pm0.03$  & $187\pm13$ & $-2.90\pm0.03$ & $-2.91\pm0.04$ \\
$42.25\pm0.03$  & $104\pm10$ & $-3.16\pm0.04$ & $-3.12\pm0.06$  \\
$42.31\pm0.03$  & $74\pm8$ & $-3.30\pm0.05$ & $-3.21\pm0.08$  \\
$42.37\pm0.03$  & $32\pm5$ & $-3.67\pm0.08$ & $-3.53\pm0.15$   \\
$42.43\pm0.03$  & $23\pm4$ & $-3.81\pm0.10$ & $-3.63\pm0.19$  \\
$42.49\pm0.03$  & $17\pm4$ & $-3.94\pm0.12$ & $-3.78\pm0.23$   \\
$42.55\pm0.03$  & $8\pm2$ & $-4.27\pm0.19$ & $-4.10\pm0.43$   \\
$42.61\pm0.03$  & $6\pm2$ & $-4.39\pm0.23$ & $-4.21\pm0.56$  \\
$42.73\pm0.03$  & $14\pm3$ & $-4.03\pm0.14$ & $-4.28\pm0.29$  \\
$42.85\pm0.09$  & $4\pm2$ & $-4.57\pm0.30$ & $-5.03\pm0.53$  \\
$43.03\pm0.09$  & $1\pm1$ & $-5.17\pm0.38$ & $-5.63\pm0.38$  \\
\end{tabular}
\label{LF_OIII_TAB}
\end{table}

%
%
%
\begin{table}
 \centering
  \caption{[O{\sc ii}] Luminosity Function not corrected for dust extinction. The total volume per bin is $26.2\times10^5$\,Mpc$^3$, assuming a top hat filter, but $ \phi$ corr has been corrected for both incompleteness and the fact that the filter profile is not a perfect top hat.}
  \begin{tabular}{@{}cccc@{}}
  \hline
  \bf $\bf \log$\bf L$_{\rm \bf [OII]}$ & \bf Sources &$\bf  \phi$ \bf obs & $\bf  \phi$ \bf corr \\
 \hline
 \bf  $\bf z=2.2$ & \# & Mpc$^{-3}$ & Mpc$^{-3}$  \\
  \hline
   \noalign{\smallskip}
$42.60\pm0.025$  & $141\pm11$ & $-2.97\pm0.04$ & $-2.88\pm0.11$  \\
$42.65\pm0.025$  & $118\pm10$ & $-3.04\pm0.04$ & $-2.96\pm0.06$  \\
$42.70\pm0.025$  & $73\pm8$ & $-3.25\pm0.05$ & $-3.14\pm0.12$ \\
$42.75\pm0.025$  & $53\pm7$ & $-3.39\pm0.06$ & $-3.31\pm0.11$  \\
$42.80\pm0.025$  & $30\pm5$ & $-3.64\pm0.09$ & $-3.54\pm0.14$  \\
$42.85\pm0.025$  & $18\pm4$ & $-3.86\pm0.12$ & $-3.75\pm0.15$ \\
$42.90\pm0.025$  & $15\pm3$ & $-3.94\pm0.13$ & $-3.84\pm0.16$ \\
$42.95\pm0.025$  & $6\pm2$ & $-4.34\pm0.23$ & $-4.21\pm0.25$   \\
$43.00\pm0.025$  & $2\pm1$ & $-4.82\pm0.53$ & $-4.69\pm0.58$  \\
$43.10\pm0.075$  & $5\pm2$ & $-4.99\pm0.26$ & $-4.99\pm0.26$ \\
$43.25\pm0.075$  & $2\pm1$ & $-5.52\pm0.53$ & $-5.52\pm0.53$  \\
\end{tabular}
\label{LF_OII_TAB}
\end{table}

\bsp

\label{lastpage}

\end{document}